\documentclass[adp,fleqn]{w-art}
\usepackage{times}
\usepackage{w-thm}
\usepackage[]{graphicx}
\begin{document}
\DOIsuffix{theDOIsuffix}
\Volume{12}
\Issue{1}
\Month{01}
\Year{2003}
\pagespan{1}{}
\Receiveddate{15 November 2002}
\Reviseddate{30 November 2002}
\Accepteddate{2 December 2002}
\Dateposted{3 December 2002}



\title[Electrodynamics of correlated electron
matter]{Electrodynamics of correlated electron matter}


\author[S.V. Dordevic]{S.V. Dordevic \inst{1,}%
  \footnote{e-mail:~\textsf{sasha@physics.uakron.edu}}}

\address[\inst{1}]{Department of Physics, The University of
Akron, Akron, OH 44325}
\author[D.N. Basov]{D.N. Basov
\inst{2,}\footnote{e-mail:~\textsf{dbasov@physics.ucsd.edu}}}

\address[\inst{2}]{Department of Physics, University of
California, San Diego, La Jolla, CA 92093}

\begin{abstract}
Infrared spectroscopy has emerged as a premier experimental
technique to probe enigmatic effects arising from strong
correlations in solids. Here we report on recent advances in this
area focusing on common patterns in correlated electron systems
including transition metal oxides, intermetallics and organic
conductors. All these materials are highly conducting substances
but their electrodynamic response is profoundly different from the
canonical Drude behavior observed in simple metals. These
unconventional properties can be attributed in several cases to
the formation of spin and/or charge ordered states, zero
temperature phase transitions and strong coupling to bosonic
modes.
\end{abstract}

\maketitle                   







\section{Introduction}
\label{intro}

Among early seminal successes of quantum mechanics is the advent
of the band theory of solids. Within the band theory it became
possible to account for fundamental distinctions between the
properties of metals and insulators without the need to invoke
interactions in the electronic systems. Transition metal oxides
including NiO and CuO are notorious examples of the failure of the
band theory. The theory prescribes a metallic state in these
compounds that are experimentally known to be insulators. Peierls
and Mott proposed that this enigma may be related to strong
Coulomb interactions between the electrons. This realization,
dating back to the late 1930-s  instituted
the field of correlated electron systems, which to this day remains
one of the most vibrant sub-areas of modern condensed matter
physics. The last two decades were devoted to the systematic
exploration of a diverse variety of materials where complex
interplay between charge, spin, lattice and orbital degrees of
freedom produces a myriad of interesting effects. These include
high-$T_c$ superconductivity, colossal magneto-resistance, spin-
and/or charge ordered states, heavy electron fluids and many
others. A common denominator between these systems is that the
interactions in the electronic systems can no longer be regarded
as "weak" as in ordinary metals or semiconductors. While standard
theories of metallic conductors are generally inadequate for
correlated electron systems it has proven difficult to develop
novel theoretical constructs elucidating the complex physics of strong
correlations.

Challenges of reaching deep understanding of correlated electron
matter prompted development and refinement of experimental
techniques most relevant for advancing the experimental picture. A
variety of spectroscopic methods traditionally played the crucial
role in establishing a description of metals and semiconductors
that currently constitutes several chapters in generic texts on
condensed matter physics. The last two decades brought
dramatic progress in spectroscopies including tunneling, inelastic
x-ray scattering, photoemission and infrared/optical spectroscopy.
In this article we attempt to discuss some of the recent successes
in the studies of the electrodynamic response of correlated
electron systems enabled by the investigation of the interaction
of the electromagnetic radiation in the $\hbar \omega$ = 1 meV -- 10
eV range in several classes of pertinent materials. The key
advantage of this particular proccedure to study correlations in
solids is that it yields information on optical constants in the
frequency region that is critical for the understanding of
elementary excitations, dynamical characteristics of
quasiparticles and collective modes. The optical constants can be
modeled using a variety of theoretical approaches. Moreover, in
the situations where the theoretical guidance to the data
interpretation is insufficient, valuable insights still can be
obtained from model independent analysis of the optical constants
using a variety of the sum rules. Equally advantageous is the
exploration of the power laws in the frequency dependence of the
optical constants that in many cases uncovers universalities
expected to occur in correlated matter in the proximity to phase
transitions.

Regardless of the complexity of a  correlated sytem response,
the Drude model of metals is usually employed as a starting point
of the examination of the electromagnetic properties. The Drude
model establishes a straightforward relationship between
electrical and optical properties of metals: a giant leap made by
Paul Drude back in 1900 \cite{drude00a,drude00b}. Indeed, a
celebrated Drude expression for the complex optical conductivity
$\tilde \sigma(\omega) = \sigma_{DC}/(1-i\omega\tau)$ gives a
specific prediction for the response of a solid at {\it all
frequencies} based on its DC conductivity $\sigma_{DC}$ and the
relaxation time $\tau$. Even though it is hard to justify the
assumptions behind the original derivation, the Drude formula has
been remarkably successful in offering simple and intuitive
descriptions of mobile carriers in metals and doped semiconductors
agreeing well with the experimental facts. As a rule, conducting
correlated electron materials reveal significant deviations from
the Drude formula. In this article we will examine physical
mechanisms underlying unconventional electromagnetic behavior of
correlated electron materials.

This article is organized in the following way. We start by
discussing experimental procedures used to obtain the optical
constants of a solid (Section \ref{exp}). We then outline the
Drude formalism and its modifications in Section \ref{drude}.
Since many correlated electron systems are derived by doping
Mott--Hubbard (MH) insulators with holes and/or electrons, we will
overview the generic features of the optical conductivity of this
class of compounds in Section \ref{doped}. Doped charges in MH
systems often form self-organized spin and/or charge ordered
states. It is instructive to explore these states in the context
of what is firmly established regarding the properties of more
conventional charge- and spin-density wave systems (Section
\ref{cdw}). In Section \ref{power} we proceed to unconventional
power laws and scaling behavior commonly encountered in correlated
electron matter. Section \ref{cuprates} presents a comprehensive
analysis of strong coupling effects in high-T$_c$ superconductors
with the goal of critically assessing some of the candidate
mechanisms of superconductivity. Section \ref{hf} focuses on the
analysis of the electrodynamics of heavy electron fluids. The
analysis of the power laws in the conductivity is continued in
Section \ref{1d} where we compare and contrast the response of
organic and inorganic 1-dimensional conductors. Concluding remarks
and outlook appear in Section \ref{summary}.


\section{Experimental observables and measurements techniques}
\label{exp}

The first step in the quantitative analysis of the electromagnetic
response of a solid is a determination of the optical constants of
a material out of experimental observables. At frequencies above
the microwave region these observables include reflectance
$R(\omega)$, transmission $T(\omega)$ and ellipsometric
coefficients $\psi(\omega)$ and $\Delta(\omega)$. As a side note,
we point out that the principles of ellipsometry were pioneered by
Paul Drude. The real and imaginary parts of complex dielectric
function $\tilde{\epsilon}(\omega)$ or the complex conductivity
$\tilde{\sigma}(\omega)$ can be inferred through one or several of
the following procedures:

\begin{enumerate}

\item A combination of reflectance $R(\omega)$ and transmission
$T(\omega)$ obtained for transparent materials can be used to
extract the dielectric function through analytic expressions. This
procedure can be used for film--on--a--substrate systems, as
described for example in Ref.~\cite{singley-03}.

\item Kramers-Kronig analysis of $R(\omega)$ for opaque systems
or of $T(\omega)$ for a transparent system.

\item  Ellipsometric coefficients for either transparent or
opaque material can be used to determine the dielectric function
through analytic expressions.

Apart from these protocols based on the intensity measurements,
two other techniques are capable of yielding the optical constants
directly through the analysis of the phase information:

\item Mach-Zehnder interferometry \cite{kozlov98,pronin98}.

\item terahertz time domain spectroscopy \cite{griskovsky90}.

\end{enumerate}

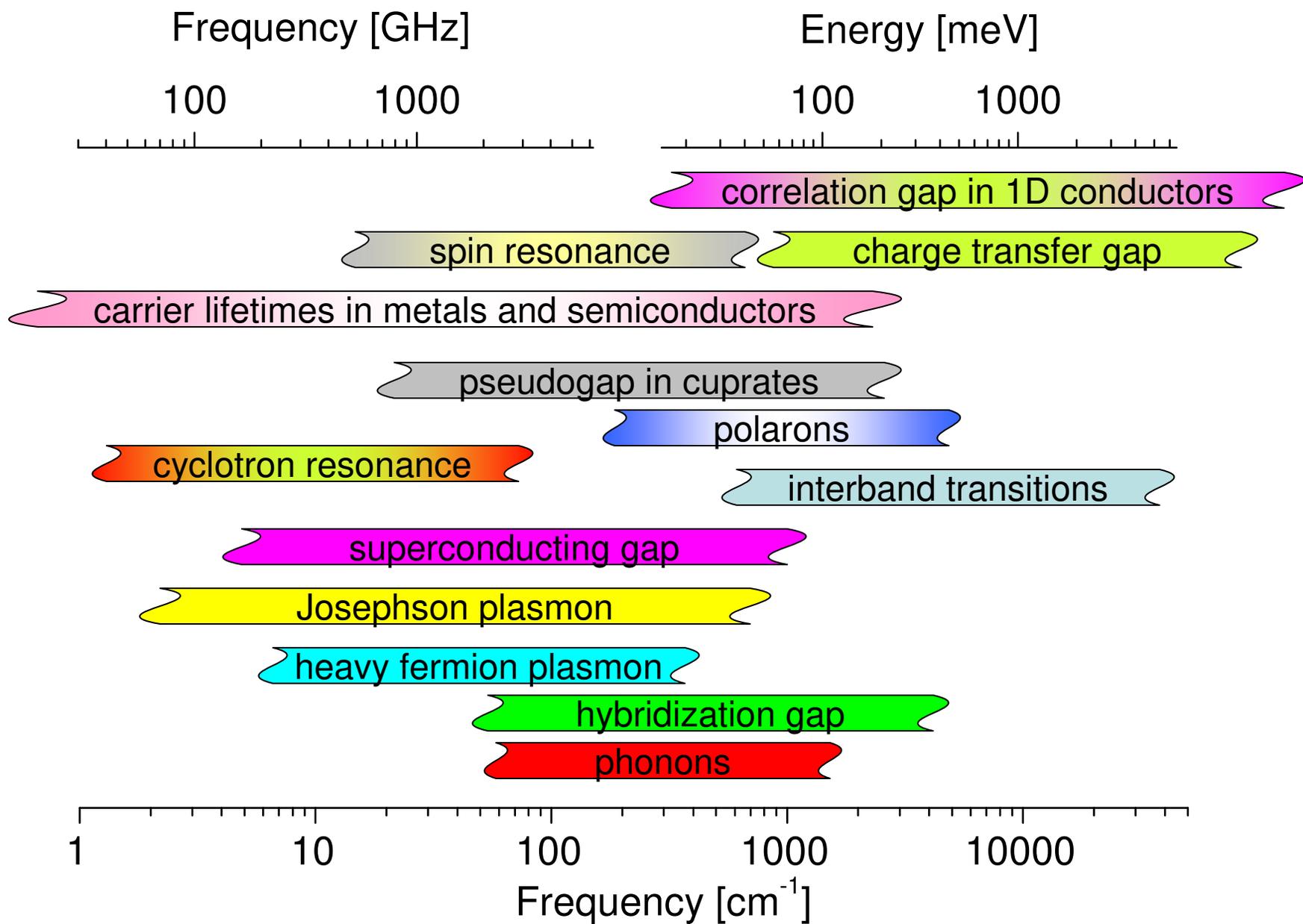
\begin{figure}[b]
\caption{The energy scales of certain phenomena in condensed
matter systems with strongly correlated electrons. Most of these
energies are in the microwave, infrared, visible and ultraviolet
parts of the spectrum. Infrared spectroscopy, combined with
microwave and optical spectrometry can be used to probe all these
phenomena simultaneously.} \label{fig:scales}
\end{figure}

All these methods have their own advantages and shortcomings. For
example, (5) is ideally suited for the survey of the temporal
evolution of the optical constants under short pulsed
photoexcitations \cite{chemla03,averitt}. However, the frequency
range accessible to this technique is inherently limited to THz
and very far-IR. It is worth noting that all these protocols
produce the optical constants throughout the entire frequency
range where experimental data exist. A notable exception is (2),
where both the low- and high-frequency extrapolations required for
KK-analysis effectively reduce the interval of reliable data.
Nevertheless, KK-analysis of reflectance is the most commonly used
technique for the extraction of the optical constants in the
current literature. Recently several groups have combined methods
(2) and (3) in order to improve the accuracy of IR measurement
over a broad frequency range \cite{burch04,kuzmenko05}.
Experimental advances allow one to carry out reflectance
measurements using micro-crystals \cite{homes93} at temperatures
down to hundreds of mK \cite{reedyk03,basov03} and in high
magnetic field \cite{padilla04}. Probably the most powerful
attribute of IR/optical spectroscopy is its ability to probe a
very broad range of photon energies, spanning several decades from
microwave to ultraviolet (Fig.~\ref{fig:scales}). This diagram
displays schematically some of the physical phenomena along with
the relevant energies. It is apparent that many of these phenomena
fall into the part of electromagnetic spectrum which can be and
has been probed using IR spectroscopy, as will be discussed below.


\section{Understanding electromagnetic response of a solid}
\label{drude}

In 1900, only 3 years after electron was discovered, P.~Drude
introduced a simple model for the DC conductivity of metals
\cite{drude00a,drude00b}:

\begin{equation}
\sigma_0=\frac{ne^2\tau}{m},
\label{eq:drudeDC}
\end{equation}
where $n$ is the carrier density, $m$ their mass and $\tau$ the
relaxation time (mean time between collisions). The model has also
been generalized for finite frequencies:

\begin{equation}
\tilde\sigma(\omega)= \sigma_1(\omega)+i\sigma_2(\omega)=
\frac{\sigma_0}{1-i \omega \tau}=\frac{1}{4 \pi}\frac{\omega_p^2
\tau}{1-i \omega \tau}, %
\label{eq:drude}
\end{equation}
where, $\omega_p^2=4 \pi e^2 n/m_b$ is the plasma frequency, $n$ is the
carrier density, m$_b$ is the carrier band mass and 1/$\tau$ is
the scattering rate. The functional form of the model predicts
that the real (dissipative) part of the optical conductivity
$\sigma_1(\omega)$ is a Lorentzian centered at $\omega$=\,0. At
higher frequencies (i.e. for $\omega \tau \gg$ 1), the model
predicts the fall-off $\sigma_1(\omega) \sim \omega^{-2}$. In
Section \ref{power} we will report on the detailed analysis of the
power law behavior of the optical conductivity of simple and
correlated metals.

In addition to this {\it intra-band} contribution, {\it
inter-band} excitations can also contribute to the optical
conductivity and they are usually modeled using finite frequency
Lorentzian oscillators:

\begin{equation}
\tilde \sigma_{lor}(\omega)= \frac{1}{4\pi}\frac{i \omega
\omega_{pj}^2}
{ \omega^{2} - \omega_{j}^{2} + i \gamma_{j}\omega},
\label{eq:lorentz}
\end{equation}
where $\omega_j$ is the frequency of the oscillator, $\gamma_j$
its width and $\omega_{pj}$ the oscillator strength. Other
excitations (such as phonons) can also contribute, and they can
also be modeled using Eq.~\ref{eq:lorentz}. A combination of
Eqs.~\ref{eq:drude} and \ref{eq:lorentz} constitutes a so-called
{\it multi-component} description to optical spectra
\cite{basov-timusk-05}.

An obvious problem with the Drude(-Lorentz) models is the
assumption of the energy independent scattering rate $1/\tau =
const$. As we will discuss below, in many real materials the
scattering rate is known to deviate strongly from this assumption.
To circumvent this limitation  of constant scattering rate in
Eq.~\ref{eq:drude}, J.W. Allen and J.C. Mikkelsen in 1977
introduced a so-called "extended Drude" model \cite{allen77} in
which the scattering rate is allowed to have frequency dependence
$1/\tau(\omega)$. Causality requires that the quasiparticle
effective mass also acquires frequency dependence
m$^*(\omega)$/m$_b$. Both quantities can be obtained from the
complex optical conductivity as:

\begin{equation}
\frac{1}{\tau(\omega)} =\frac{\omega_{p}^{2}}{4\pi}
\Re\left[\frac{1}{\tilde\sigma(\omega)}\right]=
\frac{\omega_{p}^{2}} {4 \pi} \frac{\sigma_1(\omega)}
{\sigma_1^2(\omega)+\sigma_2^2(\omega)}, %
\label{eq:tau}
\end{equation}

\begin{equation}
\frac{m^{*}(\omega)}{m_b} =\frac{\omega_{p}^{2}}{4 \pi}
\Im\left[\frac{1}{\tilde\sigma(\omega)}\right] \frac{1}{\omega} =
\frac{\omega_{p}^{2}}{4 \pi} \frac{\sigma_2(\omega)}
{\sigma_1^2(\omega)+\sigma_2^2(\omega)}\frac{1}{\omega},
\label{eq:mass}
\end{equation}
where the plasma frequency $\omega_p^2=4 \pi e^2 n/m_b$ is
estimated from the integration of $\sigma_{1}(\omega)$ up to the
frequency of the onset of interband absorption. Eqs.~\ref{eq:tau}
and \ref{eq:mass} are the basis of a so-called {\it one-component}
approach for the interpretation of optical properties
\cite{basov-timusk-05}.

The optical constants obey various sum rules
\cite{basov-timusk-05,dressel-book}. The power of these sum rules
is that they are based on the most fundamental conservation laws
and are therefore model--independent. The most famous and the most
frequently used sum rule is the one for the real part of the
optical conductivity $\sigma_1(\omega)$:

\begin{equation}
\int_{0}^{\infty}\sigma_1(\Omega)d\Omega = \frac{\omega_p^2}{8}
=\frac{\pi n e^2}{2 m_0},%
\label{eq:sum-rule}
\end{equation}
and simply expresses the conservation of charge ($n$ is the total
number of electrons in the system). Note that in
Eq.~\ref{eq:sum-rule} the integration must be performed up to
infinity in order to count all the electrons. From the practical
point of view this integration to infinity is not possible.
Instead one introduces a so-called effective spectral weight
defined as:

\begin{equation}
N_{eff}(\omega)= \frac{120}{\pi}
\int_{0+}^{\omega}\sigma_1(\Omega)d\Omega,%
\label{eq:weight}
\end{equation}
which for $\omega \rightarrow \infty$ becomes the sum rule defined
by Eq.~\ref{eq:sum-rule}. For finite integration limits the
quantity $N_{eff}(\omega)$ represents the effective number of
carriers contributing to optical absorption below the frequency
$\omega$. This quantity is particularly useful in studying
temperature- and doping--induced changes across phase transitions.


\section{Doped transition metal oxides}
\label{doped}

The last two decades have seen an explosion of interest in doped
insulators, in particular doped transition--metal oxides
\cite{imada98}. The interest stems from the fact that some of the
physical phenomena that are currently in the focus of condensed
matter research, such as high-T$_c$ superconductivity,
metal--insulator transitions and colossal magnetoresistance, are
realized in doped insulators. Infrared and optical spectroscopy
has been one of the most important experimental techniques in the
identification  of the key signatures of the electronic transport in
these novel systems. Fig.~\ref{fig:doped-insulators} shows the
in-plane optical conductivity of: La$_{1-x}$Sr$_x$TiO$_3$
\cite{fujishima-92}, La$_{1-x}$Sr$_x$VO$_3$ \cite{inaba95},
La$_{1-x}$Sr$_x$MnO$_3$ \cite{okimoto97}, La$_{1-x}$Sr$_x$CoO$_3$
\cite{tokura98}, La$_{2-x}$Sr$_x$NiO$_4$ \cite{ido91} and
La$_{2-x}$Sr$_x$CuO$_4$ \cite{uchida91}. In
Fig.~\ref{fig:doped-insulators} these compounds are arranged
according to the number of transition metal $d$-electrons, from Ti
with electron configuration [Ar]3d$^2$4s$^2$, to Cu with
configuration [Ar]3d$^{10}$4s. Note however that the actual
electronic configuration changes as La is being replaced with Sr,
as indicated in the figure. All systems shown in
Fig.~\ref{fig:doped-insulators} can be doped over a broad range of
Sr, driving the compounds through various phase transitions and
crossovers. For zero doping ($x=0$) all systems are
antiferromagnetic (AFM) Mott-Hubbard (MH) or charge transfer (CT)
insulators \cite{imada98}, as indicated in the figure. They are
all characterized by a gap in the density of states, a feature
that dominates the response of all undoped parent compounds. As
doping $x$ progresses, the gap gradually fills in at the expense
of a depression of the spectral weight associated with excitations
at higher energy. This process is also accompanied with the
development of a zero-energy Drude mode, a clear signature of
conducting carriers in the system.

Data presented in Fig.~\ref{fig:doped-insulators} uncover generic
characteristics common to diverse classes of doped MH insulators.
First, even minute changes of doping lead to radical modification
of the optical conductivity extending over the energy range beyond
several eV. Similarly, relatively small changes of temperature
(100-200\,K, i.e. 8.625--17.25\,meV) often cause dramatic effects
in the complex conductivity of doped MH systems also extending
over several eV \cite{imada98,homes93prl,Thomas,basov95-pgap}.
These aspects of the Mott transition physics are captured by
dynamical mean field theory \cite{kotliar-rmp}. The impact of
other external stimuli such as electric and magnetic field or
photo-doping on the properties of doped MH compounds has not been
systematically explored, perhaps with the exception of magnetic
fields studies of manganites \cite{tomioka96}. Another common
feature of doped MH insulators is the deviations of the free
carrier response seen in conducting phases from a simple Drude
form that will be analyzed in more details in Section \ref{power}.

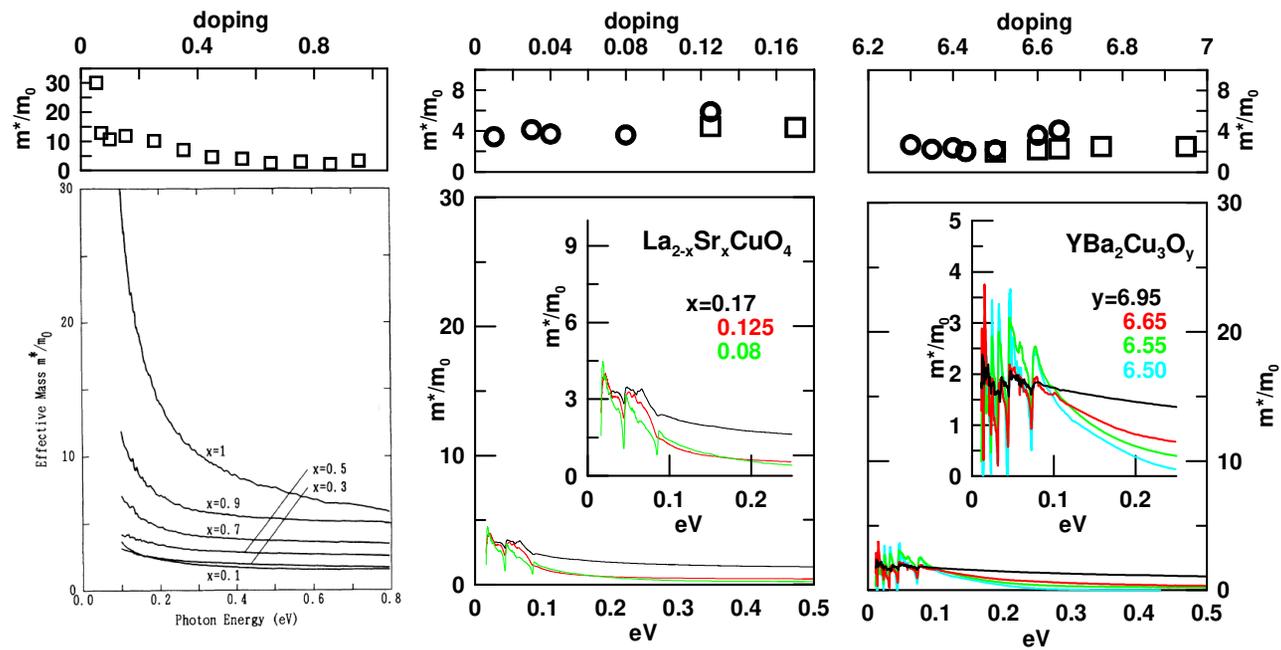
\begin{figure}[b]
\caption{Energy dependence of the carrier effective mass
$m^*(\omega)$ in Sr$_{1-x}$La$_x$TiO$_3$ (left panel)
\cite{fujishima-92}, La$_{2-x}$Sr$_x$CuO$_4$ \cite{Padilla-05} and
YBa$_2$Cu$_3$O$_y$ \cite{lee05} obtained from the extended Drude
model. The diagram shows a strong enhancement of m$^*$ values in
the titanate, but not in cuprates. The upper panels show the
doping dependence of the optical effective mass
$m*=m^*(\omega\rightarrow 0)$. The top middle and right panles
also display the optical mass values extracted from a combination
of the optical and Hall data as detailed in \cite{Padilla-05}. All
data is taken at room temperature.} \label{fig:mass}
\end{figure}

Apart from these universal trends, the dynamical characteristics
of mobile charges can be quite distinct in different materials
belonging the MH class. These differences are particularly
important as far as the behavior of the quasiparticle effective
mass m$^*$ in the vicinity of the Mott transition is concerned.
The canonical Brinkman and Rice scenario of the Mott transition
implies a divergence of m$^*$ at the transition boundary
\cite{brinkman-rice}. Imada {\it et al.} introduced the notion of
two distinct types of Mott transitions: i) those driven by
band--width and ii) those driven by carrier density
\cite{imada98}. The former type of the transition appears to be
realized in La$_{1-x}$Sr$_x$TiO$_3$ series \cite{fujishima-92},
where as high-T$_c$ superconductors belong to the latter type
\cite{Padilla-05}. In Fig.~\ref{fig:mass} we show the effective
mass spectra m$^*(\omega)$ (from Eq.~\ref{eq:mass}) for
La$_{1-x}$Sr$_x$TiO$_3$ \cite{fujishima-92} and also for
high-T$_c$ superconductors La$_{2-x}$Sr$_x$CuO$_4$ (LSCO) and
YBa$_2$Cu$_3$O$_{y}$ (YBCO) \cite{Padilla-05}. The top three
panels display the doping dependence of the estimated effective
mass m$^*$. In high-T$_c$ superconductors the effective mass is
essentially unaffected by the metal-insulator transition. On the
other hand in LSTO the mass appears to diverge as one approaches
the MH parent compound ($x=0$). The behavior of other members of
La-Sr series (Fig.~\ref{fig:doped-insulators}) remains to be
explored.


\section{Charge and spin ordered states in solids}%
\label{cdw}

Many correlated electron materials that are currently at the
forefront of condensed matter research reveal tendency toward some
form of spin and/or charge ordering at low temperatures. Ordered
states have been identified in a variety of materials, including
1D organic systems (TTF-TCNQ and (TMTSF)$_2$PF$_6$
\cite{petukhov05}), transition metal di- and tri-dichalcogenides
(NbSe$_2$ and NbS$_3$), transition metal oxides
(La$_{2-x}$Sr$_x$CuO$_4$ and La$_{2-x}$Sr$_x$NiO$_4$), curpate
ladders (Sr$_{14-x}$Ca$_x$Cu$_{24}$O$_4$), etc. In this section we
review some of these materials that have recently been studied
using infrared spectroscopy.

\subsection{Charge density wave and spin density wave systems}

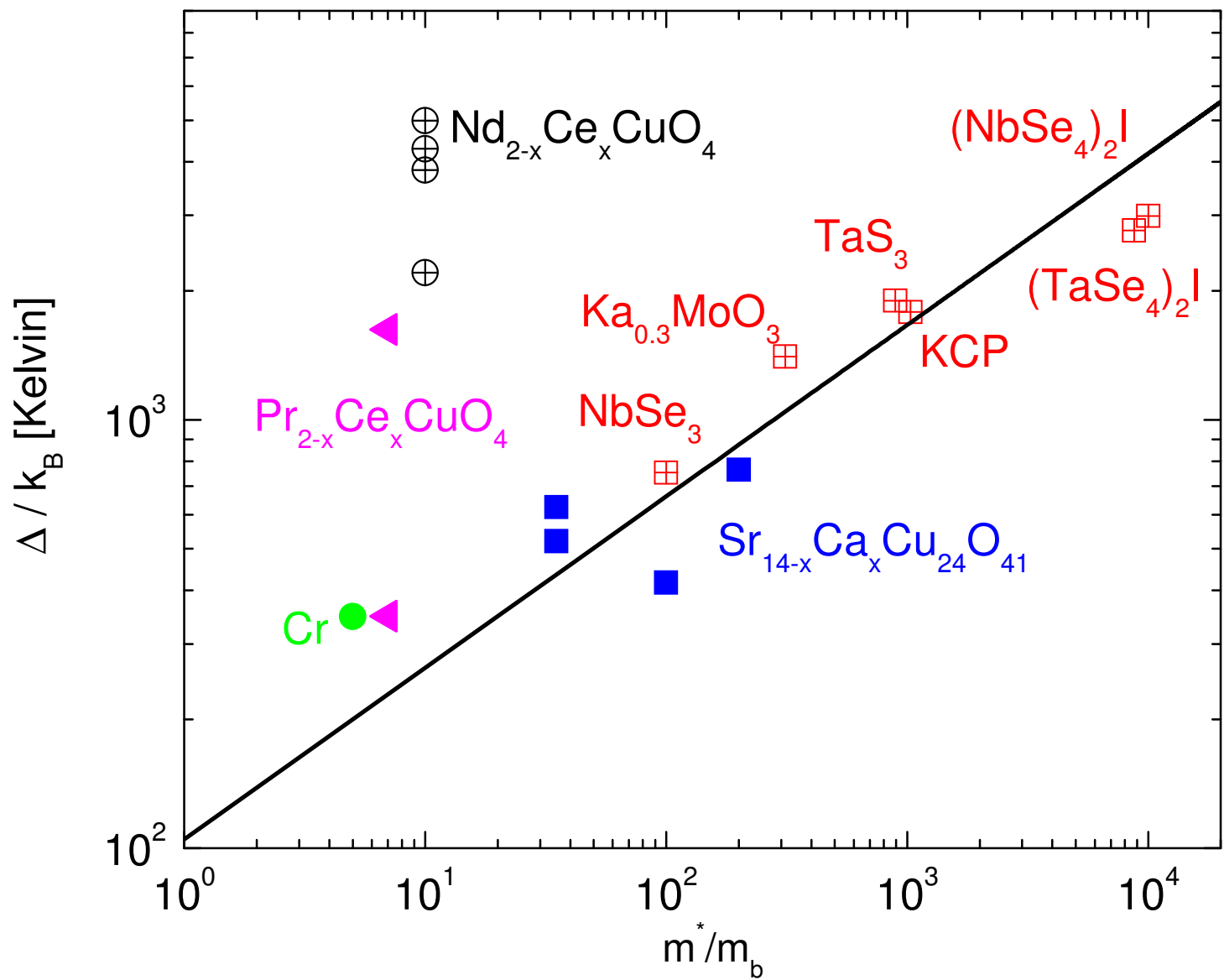
\begin{figure}[b]
\caption{Scaling between the optical gap $\Delta$ and the
effective mass m$^*$ in various CDW and SDW systems
\cite{gruner,lee74,gruner85,dressel-book}. The scaling defined by
Eq.~\ref{eq:cdw-scaling} was proposed by Lee, Rice and Anderson
\cite{lee74}. The scaling is followed by a number of
"conventional" CDW and SDW materials, but also by some transition
metal oxides with charge and/or spin ordered states.}
\label{fig:cdw-scaling}
\end{figure}

Charge density wave (CDW) and spin density wave (SDW) refer to
broken symmetry states of metals associated with the periodic
spatial variation of the electron or spin  density. Signatures of
the electromagnetic response of SDW and CDW state include two
prominent features: i) a collective mode, typically at very low
frequencies and ii) an optical gap \cite{gruner}. The optical gap
is usually described as a Bardeen--Cooper--Schrieffer (BCS) gap in
the density of states, with type-II coherence factors. One
implication of the latter coherence factors for the optical
conductivity $\sigma_1(\omega)$ is that the spectral weight from
the intragap region is transferred to frequencies just above the
gap, the total spectral weight being conserved, as demanded by the
sum rule Eq.~\ref{eq:sum-rule}. Interesting scaling between the
effective mass of the collective mode m$^*$ and the magnitude of
the single particle gap $\Delta$ was predicted theoretically by
Lee, Rice and Anderson \cite{lee74}:

\begin{equation}
\frac{m^*}{m_b}=1+\frac{4 \Delta^2}{\lambda \hbar^2
\omega_{2k_F}^2}, \label{eq:cdw-scaling}
\end{equation}
where $\lambda$ is the electron--phonon coupling constant,
$\omega_{2k_F}^2$ is the phonon frequency at 2k$_F$ and m$_b$ is
the band mass. The scaling relation $m^* \sim \Delta^2$ was
observed experimentally (Fig.~\ref{fig:cdw-scaling}) to be
followed (at least approximately) by a number of conventional CDW
systems, such as NbSe$_3$, (NbSe$_4$)$_2$I, (TaSe$_4$)$_2$I, etc.
\cite{gruner,gruner85,dressel-book}. The slope of the line in
Fig.~\ref{fig:cdw-scaling} is $\approx$\,0.4, somewhat smaller
then predicted by Eq.~\ref{eq:cdw-scaling}. In some CDW materials,
especially in 2D and 3D systems, CDW instability does not lead to
a complete gap in the density of states. For example, in quasi-2D
transition metal dichalcogenides 2H--NbSe$_2$ and 2H--TaSe$_2$,
infrared studies have shown that the CDW transition in these systems
has very little (if any) effect on the optical properties
\cite{vescoli98,dordevic01nbse,ruzicka01,dordevic03nbse}.

The formation of rigid charge density waves is also known to have
dramatic implications for optical phonons in solids: new modes are
observed corresponding to the lower symmetry of a crystal
\cite{gruner,tanner81,travaglini84,mcconnell98,homes90,minton87,jacobsen83,homes93b,creager91,calvani98,zhu02},
and also some of the resonances acquire oscillator strengths
characteristic of electronic transitions \cite{rice76}. Previous
detailed studies of IR-active phonons provided valuable insights
into the charge density wave transition in solids, as well as into
the spin Pierels physics \cite{fehske01,jones01,sushkov05}.

A canonical example of a spin density wave system is metallic
chromium Cr. Its IR spectrum has been reported by Barker {\it et
al.} \cite{barker68} and more recently by Basov {\it et al.}
\cite{basov02}. Above T$_{SDW}$=312 K the IR spectrum is typical
of metals, with a well defined Drude-like peak. A more detailed
analysis of the conductivity uncovered the frequency dependent
scattering rate 1/$\tau(\omega)$ following $\omega^2$ behavior
(Fig.~\ref{fig:cr}). The $\omega^2$ power law is expected for
canonical Fermi liquids (see Section \ref{power} below). As
temperature falls below T$_{SDW}$ a gap opens over some parts of
the Fermi surface as a consequence of SDW ordering. Gap opening
manifests itself in the IR spectra: the Drude mode narrows and the
low energy spectral weight is suppressed, resulting in a finite
frequency maximum in $\sigma_1(\omega)$ (Fig.~\ref{fig:cr}). The
optical scattering rate in the SDW state acquires a non-trivial
dependence: it is suppressed below $\omega <$\,500\,cm$^{-1}$ and
then overshoots the high-T curve. These results uncover profound
and non-intuitive consequences of an opening of a partial
(incomplete) gap in the density of states (DOS) of a metallic
system. Even though the density of states at E$_F$ is reduced, the
DC and low-$\omega$ AC conductivity are enhanced. That is because
the reduction of DOS is accompanied with the reduction of the
low-$\omega$ scattering rate. The two effects compete, and in some
cases like Cr, the latter overpowers the former, leading to "more
metallic" behavior in the gaped state. As a side remark we note
that the "area conservation" in the $1/\tau(\omega)$ data taken
above and below the SDW transition \cite{basov02}. This effect is
a property of the BCS density of states
\cite{abanov-chubukov-prl02,abanov-chubukov-basov-prb03}. The
behavior is relevant for the understanding of the pseudogap in
oxides and other correlated electron systems. Basov {\it et al.}
\cite{basov02} argued that in hole-doped cuprates the opening of a
pseudogap does not conserve the area under 1/$\tau(\omega)$
spectra. This finding is in conflict with the interpretation of
the pseudogap in terms of superconducting and/or charge/spin
density wave gap which all preserve the area in $1/\tau(\omega)$
data.

\subsection{Charge and spin ordering in strongly correlated oxides}


An unconventional form of spin and charge ordering has been
theoretically proposed and experimentally observed in transition
metal oxides, particularly nickelates and cuprates
\cite{kivelson-rmp}. In these systems strong electron-electron
correlations lead to self-organization of charge carriers into 1D
objects referred to as stripes. The role of stripes in high-$T_c$
superconductivity of cuprates is a highly debated issue
\cite{kivelson-rmp}. Unmistakable evidence for charge and spin
stripes is found for Nd-doped La$_{2-x}$Sr$_x$CuO$_4$ system. A
static stripe order is likely to induce an energy gap discussed in
the previous section. However, such a gap is not resolved in IR
measurements \cite{tajima99,dumm-prl02}. Nd-free
La$_{2-x}$Sr$_x$CuO$_4$ crystals reveal evidence for spin stripes
based on neutron scattering data whereas the corresponding charge
ordering peaks are not apparent \cite{matsuda00}. Infrared studies
carried out by Dumm {\it et al.} \cite{dumm03} uncovered
significant anisotropy of the optical conductivity with the
enhancement of $\sigma_1(\omega)$ in far-IR by as much as 50$\%$
along and the direction of the spin stripes \cite{dumm03}. Padilla
{\it et al.} searched for the evidence of phonon zone-folding
effects in La$_{2-x}$Sr$_x$CuO$_4$ crystals showing the anisotropy
of the optical conductivity attributable to spin stripes
\cite{Padilla-05}. No evidence for additional phonon peaks beyond
those predicted by group theory has been found. This latter result
points to a departure of the spin ordered state in cuprates from a
conventional stripe picture.


Some form of SDW state has been inferred from the analysis of IR
data for electron--doped cuprates, such as Pr$_{2-x}$Ce$_x$CuO$_4$
(PCCO) \cite{zimmers05} and Nd$_{2-x}$Ce$_x$CuO$_4$ (NCCO)
\cite{onose04}. Zimmers {\it et al.} have measured a series of
PCCO samples with x=0.11 (non-superconducting), 0.13 (underdoped
T$_c$=15 K), 0.15 (optimally doped T$_c$=21 K) and 0.17 (overdoped
T$_c$=15 K). In all but the overdoped sample they observed a
gap-like feature in the low-temperature spectra. Their theoretical
calculations based on tight--binding band structure augmented with
an optical gap of magnitude 0.25\,eV were able to reproduce the
most important features of $\sigma_1(\omega)$ spectra. This
optical gap was claimed to be a density wave gap induced by
commensurate ($\pi$, $\pi$) magnetic order. Interestingly the
optical gap was identified even in the optimally doped sample
(x=0.15), for which no magnetic order has been observed in neutron
scattering experiments. Similarly Onose {\it et al.} claimed SDW
in NCCO samples \cite{onose04}. Wang {\it et al.} \cite{wang04b}
reported analysis of the in-plane optical data for a series of
NCCO crystals in terms of the energy dependent scattering rate. In
underdoped materials they observe a characteristic non-monotonic
form of $1/\tau(\omega)$ consistent with the BCS form of the
density of states. These findings support the SDW interpretation
of the IR results for electron doped compounds.


Chakravarty {\it et al.} \cite{chakravarty01} proposed that the
totality of transport and spectroscopic data for holed-doped
cuprates is consistent with a new form of long--range density
wave, with d-wave symmetry of its order parameter (DDW). It has
been theoretically predicted that signatures of this peculiar
order could be observed in the infrared spectra. Valenzuela {\it
et al.} \cite{valenzuela05} and Gerami and Nayak \cite{nayak05}
have carried out numerical analysis of  possible effects in the
optical conductivity and claimed that some of these effects might
have already been seen. In particular the
transfer of spectral weight in $\sigma_1(\omega)$ with opening of
pseudogap is consistent with the predictions of the DDW model.


Recently signatures of density waves have also been identified in
cuprate ladder compounds. Osafune {\it et al.} \cite{osafune99}
investigated infrared response of the two-leg ladder system
Sr$_{14-x}$Ca$_x$Cu$_{24}$O$_4$ for two doping levels x=8 and 11.
In both cases c-axis conductivity spectra (along the ladders) at
room temperature revealed a monotonic frequency dependence.
However as temperature was lowered the optical conductivity along
c-axis $\sigma_c(\omega)$ of both compounds developed a finite
frequency peak in the far-IR part of the spectrum. Osafune {\it et
al.} argued that the peak is a collective mode of a pinned CDW, as
opposed to carrier localization that was used to explain similar
finite frequency modes in disordered cuprate superconductors
\cite{basov98}. Based on the spectral weight of the peak, Osafune
{\it et al.} estimated the effective mass to be 100 free-electron
masses for the x=\,11 sample, and 200 free-electron masses for the
x=\,8 sample. These large values signal a collective nature of the
excitations. Vuletic {\it et al.} have also studied the ladder
systems Sr$_{14-x}$Ca$_x$Cu$_{24}$O$_4$ with x=0, 3, and 9, but
inferred somewhat smaller values of the effective masses
$20<m^*<50$ \cite{vuletic03}.

Density wave states have been claimed in a number of oxide
materials which do not show static long range order in diffraction
experiments (X-ray and/or neutron scattering). For example, the
optical conductivity of a layered ruthenium oxide BaRuO$_3$
displays an opening of a pseudogap at low temperatures and
characteristic blue shift of the the spectral weight \cite{noh01}.
This was interpreted in terms of a fluctuating CDW order in the
system. Similar transfer of spectral weight was observed in
the IR spectra of ferromagnetic BaInO$_3$ \cite{cao00}. In this
quasi-1D transition metal oxide, the ferromagnetic ordering at
T$_c$=175\,K was accompanied by CDW formation. Moreover, Cao {\it
et al.} argued that the ordered magnetism was {\it driven} by CDW
formation or partial gapping of Fermi surface.

Recently discovered Na$_x$CoO$_2$ compounds attracted a lot of
attention because of the superconductivity in hydrated samples
\cite{takada03}. Sodium contents $x$ can be changed from 0 to 1,
which drives the system to various ground states. For x$>$\,3/4
Na$_x$CoO$_2$ is in a spin ordered phase, near x=2/3 it is a
Curie--Weiss metal, a charge--insulator for x$\sim$\,1/2, and a
paramagnetic metal for x$\sim$\,1/3. Infrared spectroscopy has
been performed on samples with different doping levels
\cite{lupi04,hwang04b,caimi04,wang04,bernhard04}. Samples with Na
doping x=0.5 \cite{wang04} and 0.82 \cite{bernhard04} revealed
tendency toward charge ordering. On the other hand infrared
spectra of the sample with x=0.7 doping indicated proximity to a
spin-density-wave metallic phase \cite{caimi04}.

We conclude this section with an interesting observation that the
scaling between the optical gap and effective mass
(Eq.~\ref{eq:cdw-scaling}) predicted for conventional CDW systems,
also seems to work well for cuprate ladders
Sr$_{14-x}$Ca$_x$Cu$_{24}$O$_4$ (Fig.~\ref{fig:cdw-scaling}). The
only exception is the sample with x=9, for which a very small
value for the gap was inferred from the data \cite{vuletic03}.
Osafune {\it et al.} inferred a much larger value of the gap for
similar doping levels (x=8 and 11) \cite{osafune99}. The reason for
this discrepancy is the interpretation of the far-IR feature,
which was interpreted as as a pinned CDW mode by Osafune {\it et
al.} and as the onset of a CDW gap by Vuletic {\it el al.}.
Interestingly, the scaling (Eq.~\ref{eq:cdw-scaling}) is also
followed by metallic chromium $Cr$, a canonical SDW system. On the
other hand electron-doped cuprates, such as
Pr$_{2-x}$Ce$_x$CuO$_4$ (PCCO) and Nd$_{2-x}$Ce$_x$CuO$_4$ (NCCO),
do not seem to follow the scaling.


\section{Unconventional power laws and quantum criticality}
\label{power}

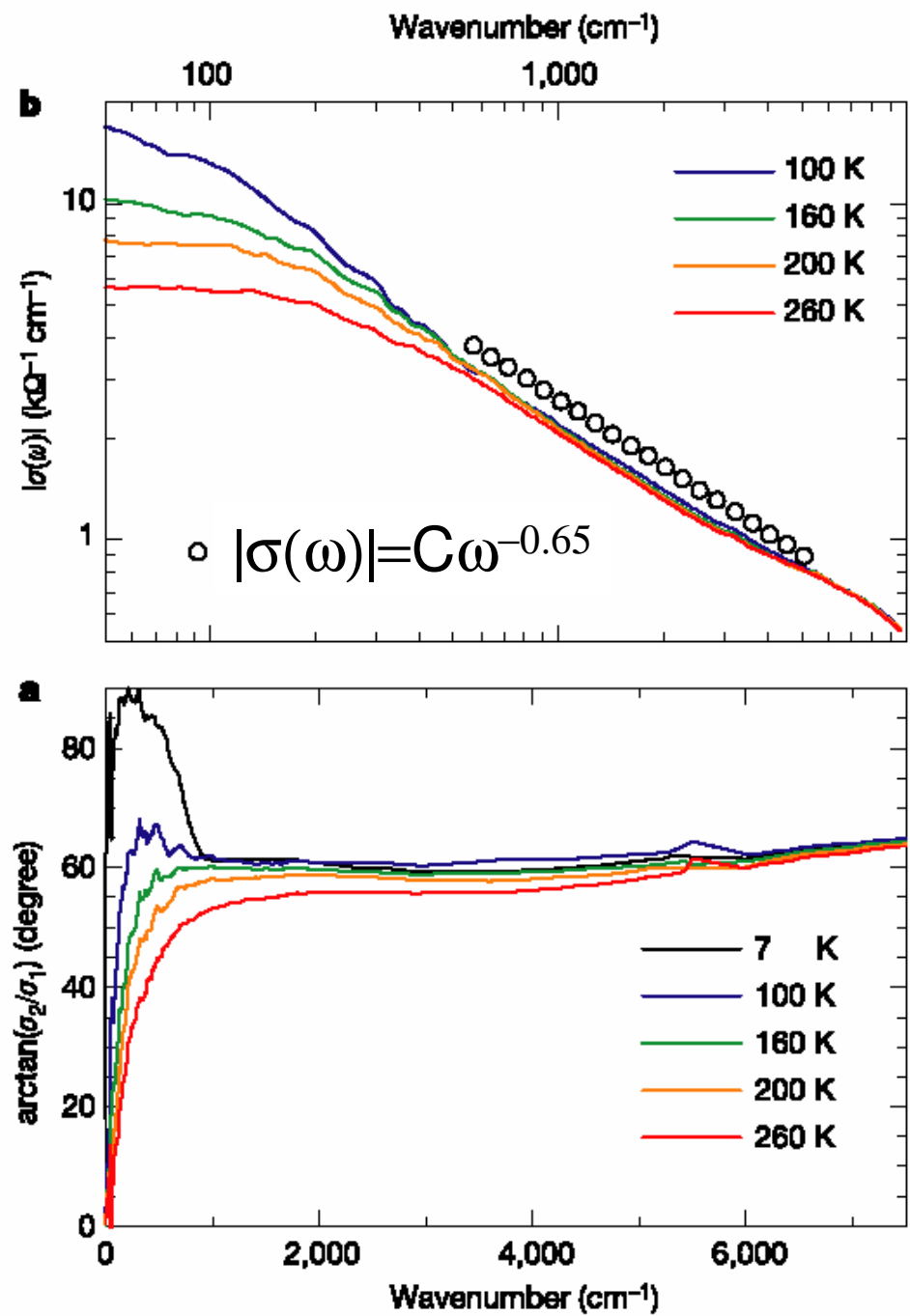
\begin{figure}[b]
\caption{Universal power laws of the optical conductivity spectra
of a nearly optimally doped high-T$_c$ superconductor
Bi$_2$Sr$_2$Ca$_{0.92}$Y$_{0.08}$Cu$_2$O$_{8-\delta}$ with
T$_c$=\,88\,K.  In (a) the absolute values of the optical
conductivity $|\sigma(\omega)|$ are plotted on a double
logarithmic scale. The open symbols correspond to the power law
$|\sigma(\omega)|=C \omega^{-0.65}$. In (b) the phase angle
function of the optical conductivity $\arctan(\sigma_2/\sigma_1)$
described in the text is presented. From van der Marel {\it et
al.} \cite{marel03}.} \label{fig:dirk1}
\end{figure}

The canonical Drude expression Eq.~\ref{eq:drude} offers an
accurate account of the conductivity for a system charges with
short-ranged interactions and predicts a power-law behavior
$\sigma_1(\omega) \propto \omega^{-2}$ at $\omega\gg1/\tau$. A
salient feature of essentially every system belonging to the class
of "strongly correlated materials" is a departure from the
$\omega^{-2}$ response. In many systems the conductivity follows
the power law form $\sigma_1(\omega) \propto \omega^{-\alpha}$
with $\alpha < 2$. In cuprate high-$T_c$ superconductors near
optimal doping, $\sigma_1(\omega)$ reveals the power law
conductivity with $\alpha\simeq$ 0.65--0.7
\cite{schlesinger90,azrak94,marel03} as exemplified in
Fig.~\ref{fig:dirk1}. The ruthenate SrRuO$_3$ $\sigma_1(\omega)$
shows  power law behavior with $\alpha\simeq 0.5$
\cite{kostic98,dodge00,yslee02}. In this Section we will present a
survey of unconventional (non-quadratic) power laws in the
conductivity of several classes of correlated electron system and
will discuss different scenarios proposed to account for this
enigmatic behavior.

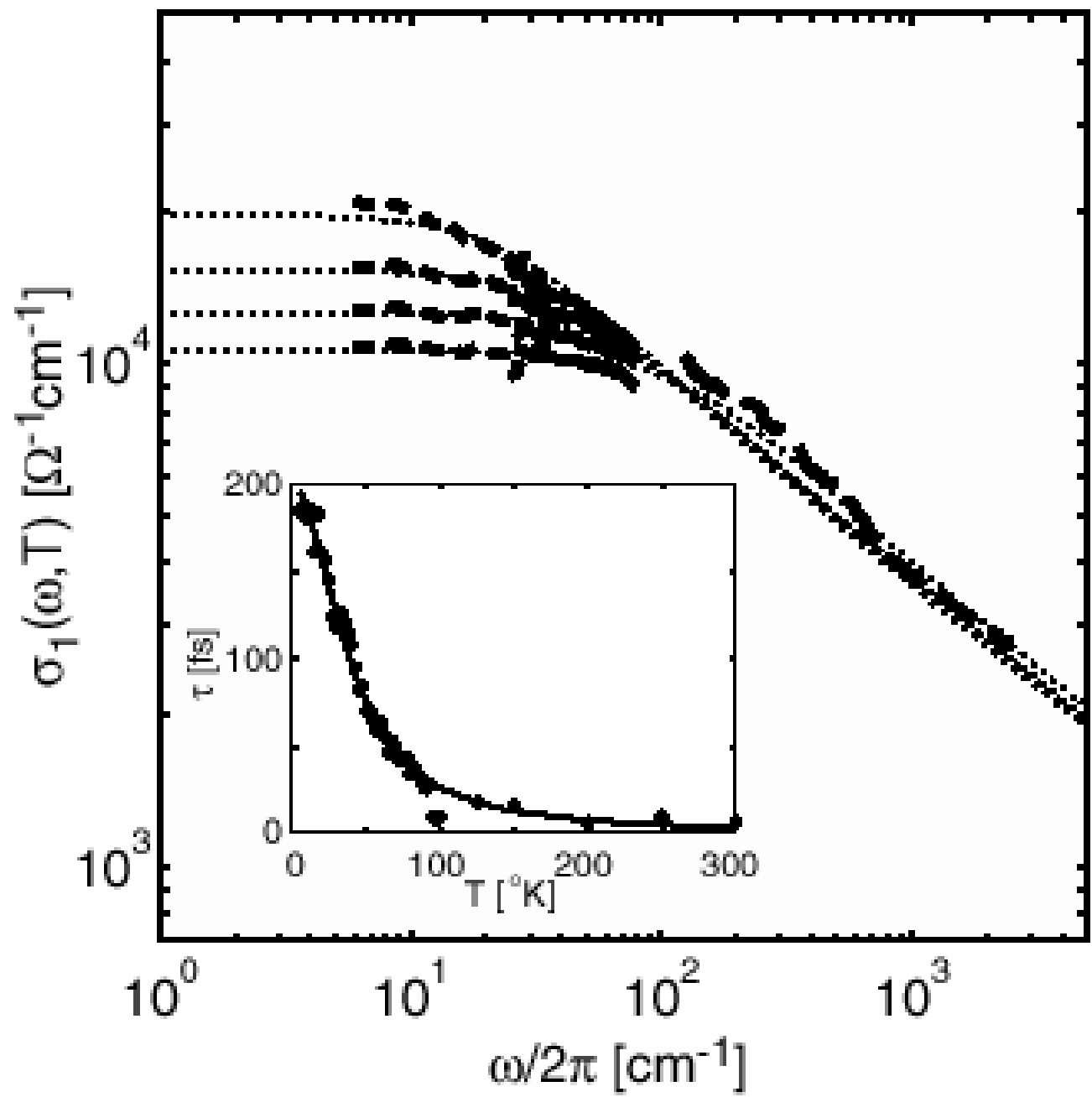
\begin{figure}[b]
\caption{Logarithmic plot of the optical conductivity of SrRuO$_3$
obtained by three methods, over three decades ranges of frequency. The
conductivity obtained from the infrared reflectivity at 40 K is
indicated by the long dashed line. Results from the far-infrared
transmission measurements, as described in the text, are indicated
by solid lines, and THz measurements by short dashed lines, with
both sets ordered in temperature from top to bottom with T= 8, 40,
60, and 80 K. Least-squared fits to
$\sigma(\omega)=A/(1/\tau-i\omega)^\alpha$ using only THz data are
shown by dotted lines. Inset: temperature dependence of $\tau$
obtained for $\alpha=0.4$ (closed circles), compared to quadratic
temperature dependence of the relaxation rate. From Dodge {\it et
al.} \cite{dodge00}.} \label{fig:dirk2}
\end{figure}

Power law scaling has been explored in great details in SrRuO$_3$
Dodge {\it et al.} \cite{dodge00} using a combination of terahertz
time domain spectroscopy by infrared reflectivity and
transmission. A borad spectral coverage has allowed the authors to
examine scaling laws over nearly three decades in frequency
(6-2400 cm$^{-1}$). These authors found the power law behavior
with $\alpha \simeq 0.4$ in the low temperature conductivity.  An
extended frequency range in the data set has uncovered unexpected
aspects of  a connection between THz/IR conductivity and the DC
transport. Spectra in Fig.~\ref{fig:dirk2} clearly show that the
divergence of $\sigma_1(\omega)$ at low frequencies is avoided on
the frequency scale set by the scattering rate $1/\tau$, thus
enforcing $\sigma_1(\omega\rightarrow 0)\propto \tau^{\alpha}$.
This observation is important since it challenges the ubiquitous
practice of inferring relaxation times from dc transport via
$\sigma_{dc}\propto \tau$ relation. Indeed, the latter relation
becomes erroneous if the conductivity follows the power law
deviating from the simple Drude result Eq.~\ref{eq:drudeDC}.
Specifically, in the case of SrRuO$_3$ the resistivity is linear
with temperature between 25 K and 120 K, implying a quadratic
temperature dependence of the relaxation rate (inset of
Fig.~\ref{fig:dirk2}). A similar non-linear relationship between
the dc data and the scattering rate has been observed in another
itinerant ferromagnet, MnSi, by Mena {\it et al.} \cite{mena03}.

It is commonly asserted that the power law behavior of the
conductivity with $\alpha  \neq 2$ signals the breakdown of the
Fermi liquid description of transport in correlated electron
systems. It is worth pointing out that in this context
$\sigma_1(\omega) \propto \omega^{-2}$ behavior is a consequence
of frequency independent scattering rate in the simple Drude
formula. However, the canonical Fermi liquid theory predicts the
scattering rate that varies quadratically both as a function of
frequency and temperature:

\begin{equation}
\frac{1}{\tau(\omega,T)} = \frac{1}{\tau_0} + a(\hbar \omega)^2 +
b(k_BT)^2 \label{eq:tau-FL}
\end{equation}
with $b/a=\pi^2$. This latter behavior is observed, for example,
in the elemental metal Cr \cite{basov02} (see Fig.~\ref{fig:cr}).
It is easy to verify that Eq.~\ref{eq:tau-FL} does not modify the
$\omega^{-2}$ character of the real part of the conductivity.
However, the imaginary part of the conductivity of a FL metal and
a Drude metal show very different power laws. In
Table~\ref{powerlaws} we display results of both analytical and
numerical calculations of optical constants $\sigma_1(\omega)$ and
$|\sigma(\omega)|=\sqrt{\sigma_1^2(\omega)+\sigma_2^2(\omega)}$
for several different models. For the models of Drude
(Eq.~\ref{eq:drude}) and Ioffe and Millis \cite{millis98}
($\sigma(\omega)=\sigma_0/\sqrt{1-i \omega \tau}$ and
$\sigma(\omega)=\sigma_0/(1-i \omega \tau)^{\alpha}$ respectively)
we derived simple analytical expressions. For Landau
(Eq.~\ref{eq:tau-FL}) and Marginal Fermi Liquid ($1/\tau(\omega)
\sim \omega$) we used KK analysis to obtain the imaginary part.

\begin{table}
\caption{The power laws of the optical constants $\sigma_1$ and
$|\sigma(\omega)|=\sqrt{\sigma_1^2(\omega)+\sigma_2^2(\omega)}$.
For Drude, and Ioffe--Millis' models analytical calculations were
employed, whereas for the Marginal Fermi Liquid and Landau Fermi
Liquid we used simple numerical calculations.}
\label{powerlaws}%
\centering
\begin{tabular}{|c|c|c|}
 \hline
& $\sigma_1(\omega)$ & $|\sigma(\omega)|$ \\
    \hline
  Drude & $\omega^{-2}$ & $\omega^{-1}$  \\
  Landau FL & $\omega^{-2}$ & $\omega^{-2}$  \\
  Marginal FL & $\omega^{-0.45}$ & $\omega^{-0.725}$  \\
 Ioffe--Millis & $\omega^{-0.5}$ & $\omega^{-0.5}$  \\
 Generalized I-M & $\omega^{-\alpha}$ & $\omega^{-\alpha}$  \\
   \hline
\end{tabular}
\end{table}

We now discuss several theoretical proposals to account for the
unusual power laws both within the Fermi liquid theory and also
using non-Fermi Liquid approaches. In order to account for the
non-Drude power law behavior of the conductivity within the FL
approaches one is forced to invoke rather anomalous scattering
mechanisms. One of the earliest accounts is the so-called Marginal
Fermi Liquid (MFL) theory postulating scattering of quasiparticles
from a bosonic spectrum that is flat over frequency scale from
$T<\omega<\omega_c$ where $\omega_c$ is a cut-off frequency
\cite{varma91}. The MFL theory assumes the following form of the
electronic self-energy $\Sigma(\omega)=\lambda[\omega
ln(x/\omega_c)+i\pi x]$ where x=$max(\omega,T)$. The model is in
fair agreement with experiments on high-$T_c$ cuprates especially
at not so low temperatures \cite{hwang04}. Alternatively, power
law behavior of the optical constants has been explored assuming
strong momentum dependence of the quasiparticle lifetime along the
Fermi surface \cite{carrington92,stojkovic96,hlubina95,millis98}.
Interestingly, this momentum dependence originally inferred from
the analysis of transport and infrared data in
Refs.~\cite{carrington92,stojkovic96,millis98} was later confirmed
by direct measurements using angle resolved photoemission
spectroscopy (ARPES) \cite{dalmacelli03}. Specifically, the "cold
spots" model of Ioffe and Millis not only predicts the power law
behavior of the conductivity with $\alpha\simeq 0.5$ for the
in-plane conductivity, but also offers an account of the gross
features of the interplane transport \cite{millis98,ioffe99}. The
momentum dependence of the quasiparticle lifetime is a natural
feature of the spin fluctuation model of Chubukov {\it et al.}
\cite{chubukov03}. The power law behavior of the conductivity
within this model has been analyzed in Ref.~\cite{abanov03} in the
context of crossover from FL to non-FL behavior above a
characteristic energy $\omega_{sf}$. Relevance of spin
fluctuations to the power law behavior of the optical data with
$\alpha\simeq 0.65$ also emerges out of numerical results obtained
within the t-J model by Zemljic and Prelovsek \cite{zemljic05}.

Unlike the (generalized) Fermi liquid models, the non-FL
approaches propose that current carrying objects are not simple
electrons but instead may involve spinons and holons
\cite{anderson87} or phase fluctuations of the superconducting
order parameter \cite{emery}. Both of these approaches predict the
power law behavior of the optical constants. Ioffe and Kotliar
considered optical conductivity of spin-charge separated systems
and derived the form $\sigma_1(\omega)\propto 1/(T+\omega^\alpha)$
\cite{ioffe90,millis98}. Depending whether spinons or holons
dominate the transport the magnitude of $\alpha$ is predicted to
be 4/3 or 3/2. Anderson proposed that a non-Drude power law is a
natural consequence of the Luttinger-liquid theory with
spin-charge separation \cite{anderson97}.

The analysis of the power law behavior and of the
frequency/temperature scaling of the optical constants is
interesting in the context of quantum phase transitions occurring
at $T\rightarrow 0$ to the electronic properties of correlated
electron systems. In the vicinity of such a transition a response
of a system to external stimuli is expected to follow universal
trends defined by the quantum mechanical nature of fluctuations
\cite{sondhi-rmp97,sachdev-book,coleman-schofield}. Van der Marel
observed $\omega/T$ scaling in the response of optimally doped
high-T$_c$ superconductor
Bi$_2$Sr$_2$Ca$_{0.92}$Y$_{0.08}$Cu$_2$O$_{8-\delta}$
\cite{marel03}. Both under- and over-doped counterparts of this
system did not reveal a similar scaling. This result is indicative
of the association of the quantum critical with optimized
transition temperature in high-T$_c$ materials. Lee {\it et al.}
reported an observation of $\omega/T$ scaling of the optical
conductivity in CaRuO$_3$ system \cite{yslee02}. These authors
attributed the quantum critical point to a zero temperature
transition between ferromagnetic and paramagnetic phases.

We conclude this section by pointing out another class of
materials with "non-FL" electrodynamics. Degiorgi {\it et al.}
investigated a variety of heavy-electron systems which they also
qualified as "non-Fermi-Liquid" \cite{degiorgi-97,degiorgi99}.
These materials (such as: U$_{0.2}$Y$_{0.8}$Pd$_3$,
UCu$_{3.5}$Pd$_{1.5}$, U$_{1-x}$Th$_x$Pd$_2$Al$_3$, and others)
typically show a non-monotonic form of the conductivity dominated
by a stark resonance at mid-IR frequencies. This non--monotonic
form of $\sigma_1(\omega)$ distinguishes U-based non--FL compounds
from doped MH insulators discussed in this sections. The non-FL
behaviour in these systems is usually explained in terms of
multichannel Kondo models, models based on proximity to quantum
critical points, or models based on a disorder \cite{stewart01}.


\section{Strong coupling effects in cuprates}
\label{cuprates}

It is widely believed that charge carries in cuprates are strongly
coupled to bosonic excitations. Signatures of strong coupling
have been evidenced by a number of experimental techniques such as
IR, ARPES and tunneling. However the nature of bosonic excitations
to which carriers are coupled is currently one of the most debated
subject in the field. Both phonons
\cite{bogdanov00,lanzara01,shen03,zhou03} and spin fluctuations
\cite{pines90,carbotte99,munzar99} are currently considered as
possible candidates. At issue is whether or not the magnetic mode
is capable of having a serious impact on the electronic
self-energy, in view of the small intensity of the resonance
\cite{abanov02,kee02,chubukov04}.

In order to discriminate between the candidate mechanisms it is
desirable to learn as many details as possible regarding the
Eliashberg spectral function $\alpha^2F(\omega)$ quantifying
strong coupling effects. This is a challenging task. The key
complication is that physical processes unrelated to strong
coupling can in principle mimic spectral signatures of
quasiparticles coupled to bosonic modes. Specifically, the energy
gap or pseudogap is known to produce a characteristic threshold
structure in the $1/\tau(\omega)$ data
\cite{Basov-96,puchkov96,puchkov96prl} that makes it difficult to
infer $\alpha^2F(\omega)$ from the raw data. It is therefore
imperative to treat consistently the two effects: (pseudo)gaps in
the density of states and the coupling of charge carriers
to some bosonic mode \cite{puchkov96}. We will discuss a new
approach to this problem recently proposed by Dordevic {\it et
al.} \cite{dordevic05} after giving the necessary background.

In 1971 P. Allen studied signatures of strong electron-phonon
coupling in the infrared spectra of metals and derived the
following formula for the optical scattering rate at T=\,0\,K
\cite{allen71}:

\begin{equation}
\frac{1}{\tau(\omega)}=\frac{2\pi}{\omega}
\int_0^{\omega}d\Omega(\omega-\Omega)\alpha^2F(\Omega).
\label{eq:tau1}
\end{equation}
This T=0 results has been generalized to finite temperatures by
Millis {\it et al.} \cite{millis88} and Shulga {\it et al.}
\cite{shulga91}:

\begin{equation}
\frac{1}{\tau(\omega,T)}=\frac{\pi}{\omega}
\int_0^{\infty}d\Omega\alpha^2F(\Omega,T) \Big[2\omega
\coth\Big(\frac{\Omega}{2T}\Big)-
(\omega+\Omega)\coth\Big(\frac{\omega+\Omega}{2T}\Big)+(\omega-
\Omega)
\coth\Big(\frac{\omega-\Omega}{2T}\Big)\Big], \label{eq:tau2}
\end{equation}
which in the limit T$\rightarrow$\,0\,K reduces to Allen's result
Eq.~(\ref{eq:tau1}).

The integral form of Eqs.~\ref{eq:tau1} and \ref{eq:tau2} implies
that the task of extraction of $\alpha^2F(\omega)$ from
experimental data is non--straightforward and calls for the
development of suitable inversion protocols. To extract the
electron-phonon spectral function in a conventional superconductor
lead Pb Marsiglio {\it et al.} employed the differential form of
Eq.~\ref{eq:tau1} \cite{marsiglio98}:

\begin{equation}
W(\omega)=\frac{1}{2\pi}\frac{d^2}{d\omega^2} \left[\omega \cdot
\frac{1}{\tau(\omega )} \right], \label{eq:w1}
\end{equation}
where W($\omega$) is a function closely related to
$\alpha^2F(\omega)$ \cite{marsiglio98}. The spectral function
obtained with Eq.~\ref{eq:w1} was in good agreement with
$\alpha^2F(\omega)$ inferred from tunneling measurements
\cite{marsiglio98}. Eq.~\ref{eq:w1} requires taking the second
derivative of experimental data, which introduces significant
numerical difficulties. The experimental data must be smoothed
before derivatives can be taken. In order to avoid the
differentiation problems Tu {\it et al.} \cite{tu02} and Wang {\it
et al.} \cite{wang03} fitted the optical spectra (reflectance or
scattering rate) with polynomials and then used Eq.~\ref{eq:w1} to
calculate $\alpha^2F(\omega)$. Casek {\it et al.} \cite{casek05}
have also used Eq.~\ref{eq:w1} to generate the electron-boson
spectral function from the model conductivity spectra, calculated
within the spin-fermion model. Hwang {\it et al.} have modeled the
scattering rate spectra 1/$\tau(\omega)$ with the formula which
includes a non-constant density of states around the Fermi level
\cite{sharapov05,hwang05}.

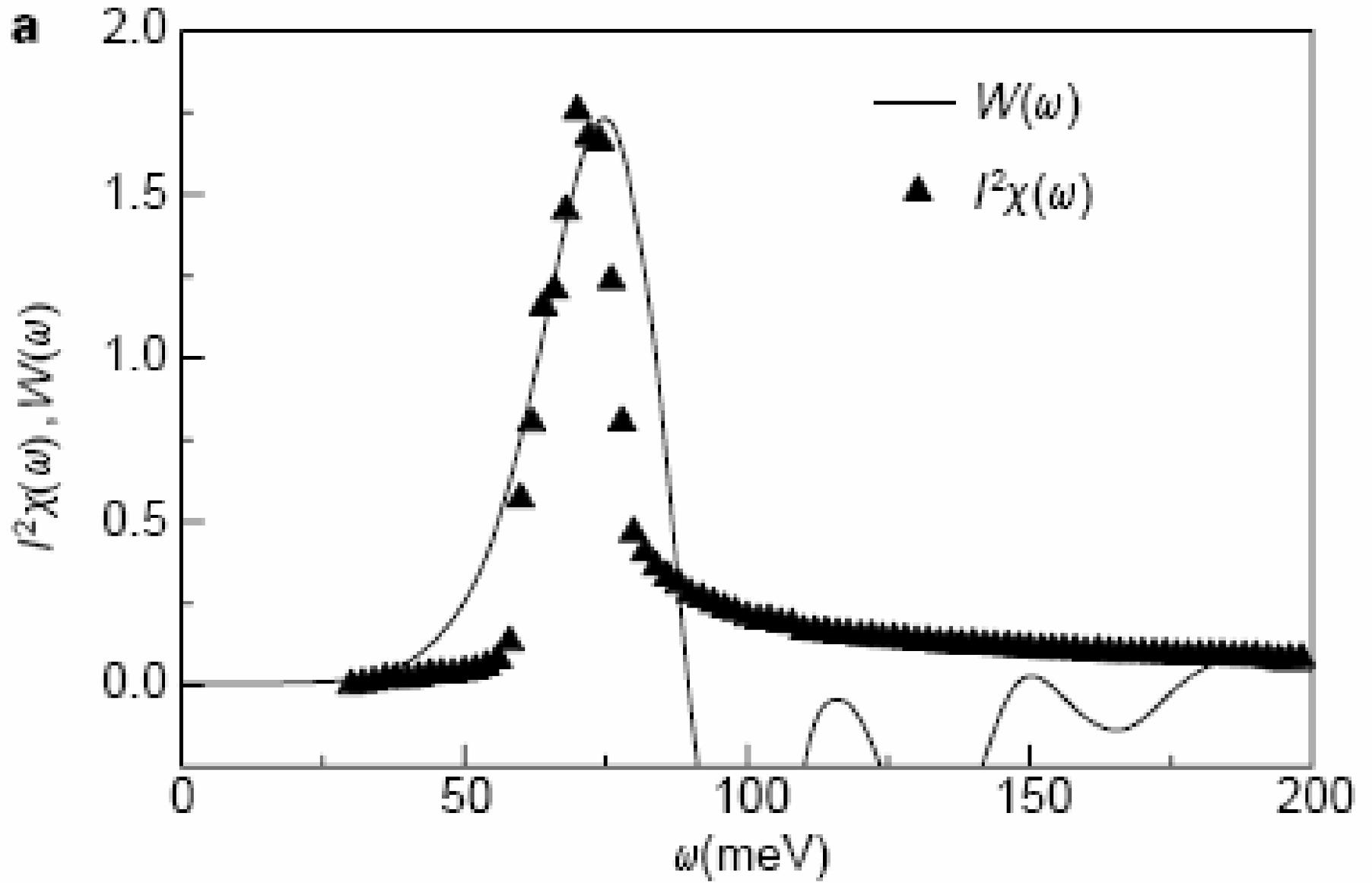
\begin{figure}[b]

\caption{The electron-boson spectral function $W(\omega)$ for
optimally doped cuprate YBa$_2$Cu$_3$O$_{6.95}$ (full line) and
the spin excitation spectrum $I^2\chi(\omega)$ (full symbols)
\cite{carbotte99}. The $W(\omega)$ spectrum was obtained using
Eq.~\ref{eq:w1} and $I^2\chi(\omega)$ is from neutron scattering
experiments. Eq.~\ref{eq:w1} does not take into account a gap in
the density of states, which results in unphysical negative values
in the spectral function. From Carbotte {\it et al.}
\cite{carbotte99}.} \label{fig:ybco}
\end{figure}

Carbotte, Schachinger and Basov have also applied Eq.~\ref{eq:w1}
to extract the electron-boson spectral function in the high-T$_c$
cuprates \cite{carbotte99}. Fig.~\ref{fig:ybco} displays an
example of these calculations for optimally doped
YBa$_2$Cu$_3$O$_{6.95}$. The electron-boson spectral function
$\alpha^2F(\omega)$ extracted this way shows a characteristic
shape, dominated by a strong peak, followed by a negative dip.
Carbotte {\it et al.} interpreted the peak as due to coupling of
charge carriers to neutron "resonance" \cite{carbotte99}.
Previously, the idea of quasipartcles coupling to a neutron mode
had been proposed by Norman and Ding \cite{norman98} and Munzar
\cite{munzar99}.

A cursory examination of Fig.~\ref{fig:ybco} uncovers one obvious
problem of the second-derivative protocol: negative values in
$\alpha^2F(\omega)$ are unphysical. This artifact of the analysis
stems from either the numerical instabilities or the application
of Eqs.~\ref{eq:tau1} or \ref{eq:w1} to systems with a (pseudo)gap
in the density of states \cite{chubukov-00}. In his celebrated
1971 article \cite{allen71} P.~Allen also derived the formula for
the scattering rate in the superconducting state at T=0\,K, i.e.
in the presence of a gap in the density of states:

\begin{equation}
\frac{1}{\tau(\omega)}=\frac{2\pi}{\omega}\int_{0}^{\omega-
2\Delta} d\Omega (\omega-\Omega) \alpha^2F(\Omega) E \Big[
\sqrt{1-\frac{4 \Delta^2}{(\omega-\Omega)^2} } \Big]
\label{eq:bcs-tau}
\end{equation}
where $\Delta$ is the (momentum-independent) energy gap and $E(x)$
is the complete elliptic integral of the second kind. For
$\Delta$=0 Eq.~\ref{eq:bcs-tau} reduces to Eq.~\ref{eq:tau1}.

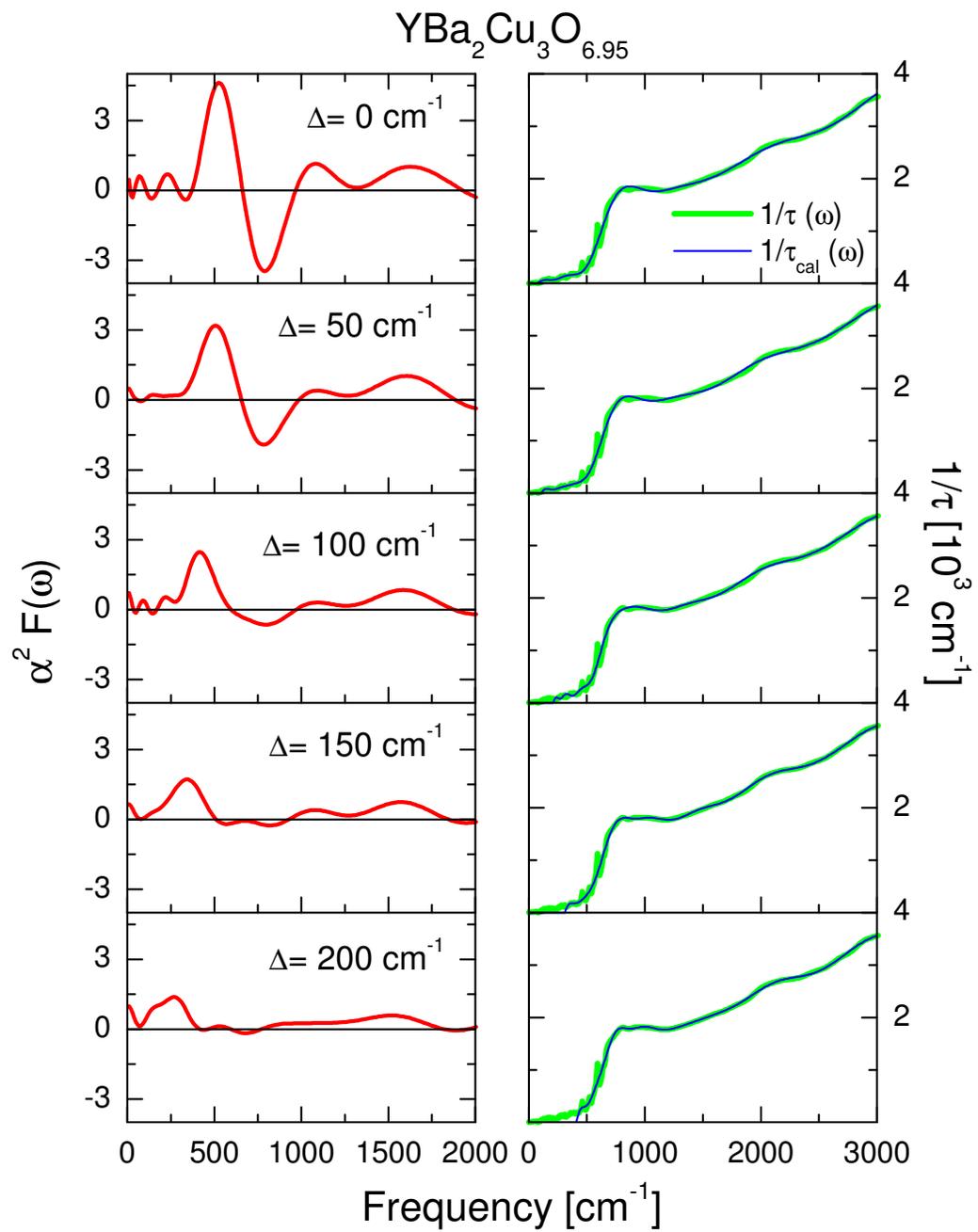
\begin{figure}[b]
\caption{The inversion calculations of the electron-boson spectral
function $\alpha^2F(\omega)$ with the formula with the gap
Eq.~\ref{eq:bcs-tau} for optimally doped YBa$_2$Cu$_3$O$_{6.95}$
(the same as in Fig.~\ref{fig:ybco}) \cite{dordevic05}. The left
panels display $\alpha^2F(\omega)$ and the right panels scattering
rate data 1/$\tau(\omega)$ and 1/$\tau_{cal}(\omega)$ calculated
from the spectral function on the left. For $\Delta=0$ the result
displays strong negative deep, but as $\Delta$ increases, the deep
is progressively suppressed. These calculations illustrate that
the origin of the negative value in the spectral function is
directly related to the gap in the DOS. Negative values can be
eliminated when the appropriate energy gap is used.}
\label{fig:ybco-gap}
\end{figure}

The extraction of $\alpha^2F(\omega)$ from Eq.~\ref{eq:bcs-tau} is
non-trivial, as the corresponding differential expression does not
exist. The integral equation must be solved (inverted) directly.
Dordevic {\it et al.} have used a so-called singular value
decomposition (SVD) method to calculate the spectral function from
Eq.~\ref{eq:bcs-tau} \cite{dordevic05}. Fig.~\ref{fig:ybco-gap}
displays inversion calculations for optimally doped YBCO (the same
as in Fig.~\ref{fig:ybco}) with several different gap values

$\Delta$. As the gap value increases, the negative dip in the
$\alpha^2F(\omega)$ spectra diminishes and for $\Delta
\approx$\,150\,cm$^{-1}$ it is almost completely absent. An
obvious problem with these calculations is that they assume: i) s-wave
gap and ii) T=\,0\ limit. Neither of these assumptions are valid for
cuprates. However even with these inadequacies
Eq.~\ref{eq:bcs-tau} is useful, because it allows one to treat
both effects (coupling to bosonic mode and gap in the density of
states) on equal footing. Carbotte and Schachinger have developed
similar analysis protocol which takes into account d-wave symmetry
of the order parameter \cite{carbotte05}.

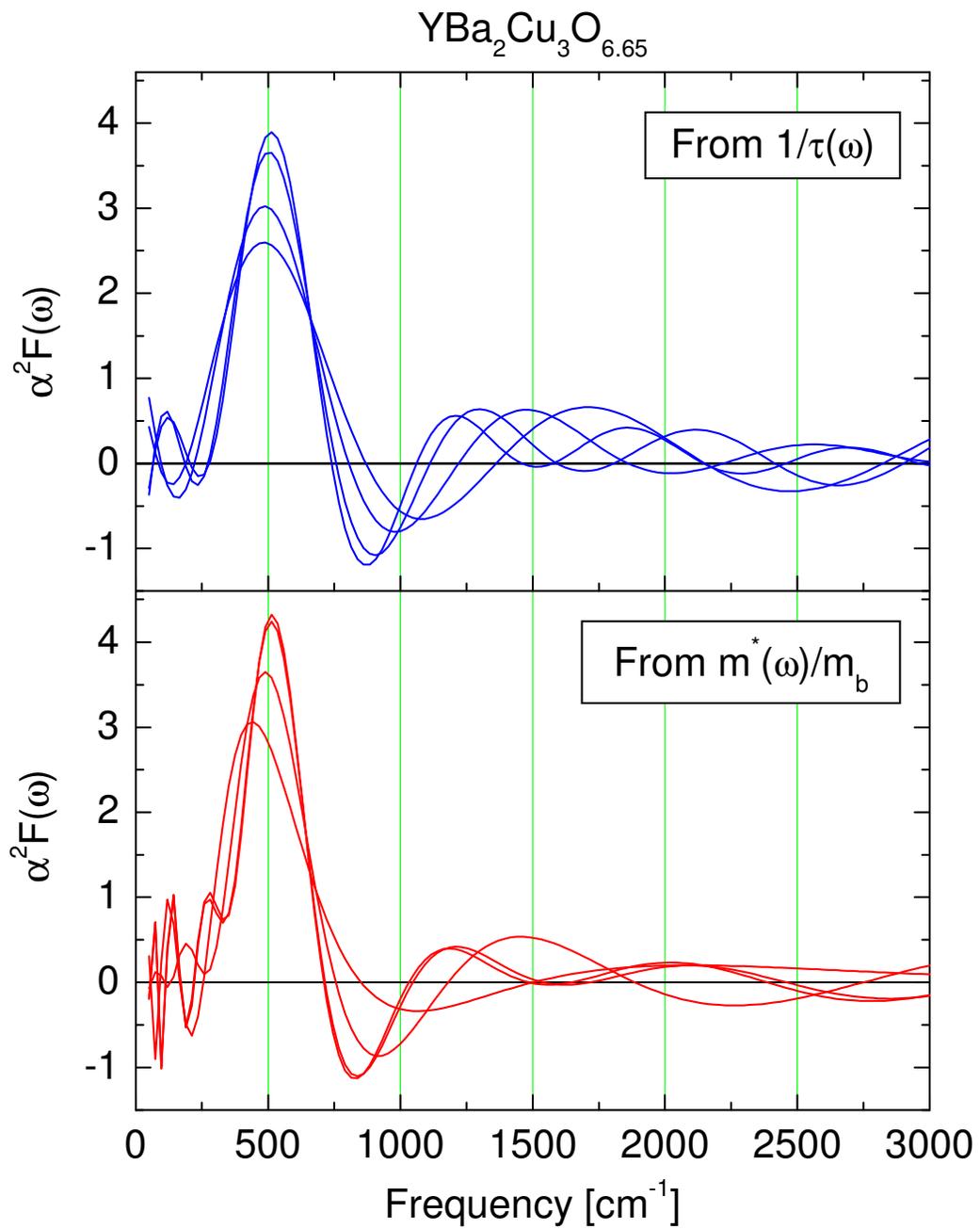
\begin{figure}[b]
\caption{The electron-boson spectral function $\alpha^2F(\omega)$
for underdoped high-T$_c$ superconductor YBa$_2$Cu$_3$O$_{6.65}$
with T$_c\simeq$\,60\,K. The top panel display $\alpha^2F(\omega)$
calculated from the scattering rate 1/$\tau(\omega)$ using
Eqs.~\ref{eq:tau1} or \ref{eq:tau2} \cite{dordevic05}. The bottom
panel shows $\alpha^2F(\omega)$ calculated from the corresponding
effective mass $m^*(\omega)/m_b$ spectra using
Eq.~\ref{eq:andrey}. In each panel, different curves represent
calculations with different number of singular values, as
explained in \cite{dordevic05}. Eqs.~\ref{eq:tau1}, \ref{eq:tau2}
and \ref{eq:andrey} do not take into account a gap in the density
of states, which results in unphysical negative values.}
\label{fig:mass-inversion}
\end{figure}

The extraction of the spectral function $\alpha^2F(\omega)$ from
$1/\tau(\omega)$ spectra can be complemented with the analogous
inversion procedure applied to the effective mass spectra
m$^*(\omega)$/m$_b$ (Eq.~\ref{eq:mass}). The corresponding formula
derived by Chubukov \cite{chubukov}:

\begin{equation}
\frac{m^*(\omega)}{m_0}=1+2\int_0^\infty\alpha^2F(\Omega) \Big[
\frac{1}{\omega}\log\Big|\frac{\Omega+\omega}{\Omega-\omega}\Big|
+\frac{\Omega}{\omega^2} \log\Big|\frac{\Omega^2-
\omega^2}{\Omega^2}\Big| \Big]d \Omega, \label{eq:andrey}
\end{equation}
is mathematically of the same type as Eqs.~\ref{eq:tau2} and
\ref{eq:bcs-tau} (the so-called Fredholm integral equation of the
first kind). Figure \ref{fig:mass-inversion} displays results of
numerical calculations of the spectral function of underdoped
YBa$_2$Cu$_3$O$_{6.65}$. The top panel is $\alpha^2F(\omega)$
obtained from the scattering rate 1/$\tau(\omega)$ (using
Eq.~\ref{eq:tau2}) and the bottom panel from effective mass
$m^*(\omega)/m_b$ (using Eq.~\ref{eq:andrey}). Both inversions
produce similar results, in particular a strong peak, followed by
a negative dip.

The inversion protocol based on the SVD algorithm can also be
applied to ARPES data \cite{dordevic05}. The procedure is based on
the analysis of the real part of the electron self-energy
($\Sigma_1(\omega)$), which is experimentally accessible through
ARPES. One must keep in mind that the spectral functions probed in
The two experiments are not the same: ARPES probes the equilibrium
$\alpha^2F(\omega)$ which is a single-particle property, whereas
IR measures transport $\alpha_{tr}^2F(\omega)$, a two-particle
property. Moreover ARPES is a momentum resolving technique,
whereas IR averages over the Fermi surface. In the latter case,
the effect of a d-wave energy gap must be taken into account,
which effectively shifts the peak up to higher energies. In the
ARPES case the energy gap (either superconducitng or pseudogap)
should not play a role, if the data is taken along nodal
direction.

The spectral functions obtained from ARPES have some similarities,
but also some important differences compared with IR. The
resolution of ARPES data is presently poorer then IR, which
implies that even fewer features could be resolved in the spectral
function. The ARPES $\alpha^2F(\omega)$ also has a strong peak,
but unlike IR the negative dip seems to be absent. Also the
high-energy contribution is not identified in the ARPES results.
The position of the main peak is particularly interesting. The
most direct comparison can be made for optimally doped
Bi$_2$Sr$_2$CaCu$_2$O$_{8+\delta}$ (Bi2212), for which both high
quality IR and ARPES data exist. The analysis has shown that the
peaks actually occur at roughly the same frequency in both ARPES
and IR spectral functions. This result is surprising and
unexpected, as the main peak in the IR $\alpha^2F(\omega)$ is
supposed to occur at higher frequencies (shifted by the gap).
Hwang {\it et al.} have made a similar comparison between the
electron self-energy obtained from IR and ARPES for several
different doping levels \cite{hwang04nature}. High quality ARPES
and IR data for the cuprate families of LSCO and NCCO have
recently become available
\cite{Padilla-05,zimmers05,zhou03,armitage03} and future analysis
will reveal if the behavior observed in Bi2212 is universal.


\section{Heavy fermion systems}
\label{hf}

Heavy fermion metals are intermetallic compounds which contain
certain rare-earth elements, such as U, Ce, or Yb. They are
characterized by large enhancements of their quaisparticle
effective mass m$^*$. The electronic properties of HF metals are
in accord with a scenario based on hybridization of localized $f$
electrons and delocalized $s$, $p$ or $d$ electrons
\cite{kotliar-rmp,millis87,millis87b,coleman87,fulde93,hewson97}.
At high temperatures the hybridization is negligible and the
properties are described by two independent sets of electrons.
However, at low temperatures hybridization leads to mixing of
conduction and f-electrons and opening of a gap in the density of
states: a so-called hybridization gap. Infrared spectroscopy is
ideally suited to probe the electronic processes in heavy fermion
systems. All hallmarks of the HF state can be simultaneously
probed using this technique. In particular, the Drude-like mode
associated with the response of heavy quasiparticles as well as
excitations across the hybridization gap have been found and
identified as characteristic signatures of the HF state in the IR
spectra \cite{degiorgi99,dordevic01}.

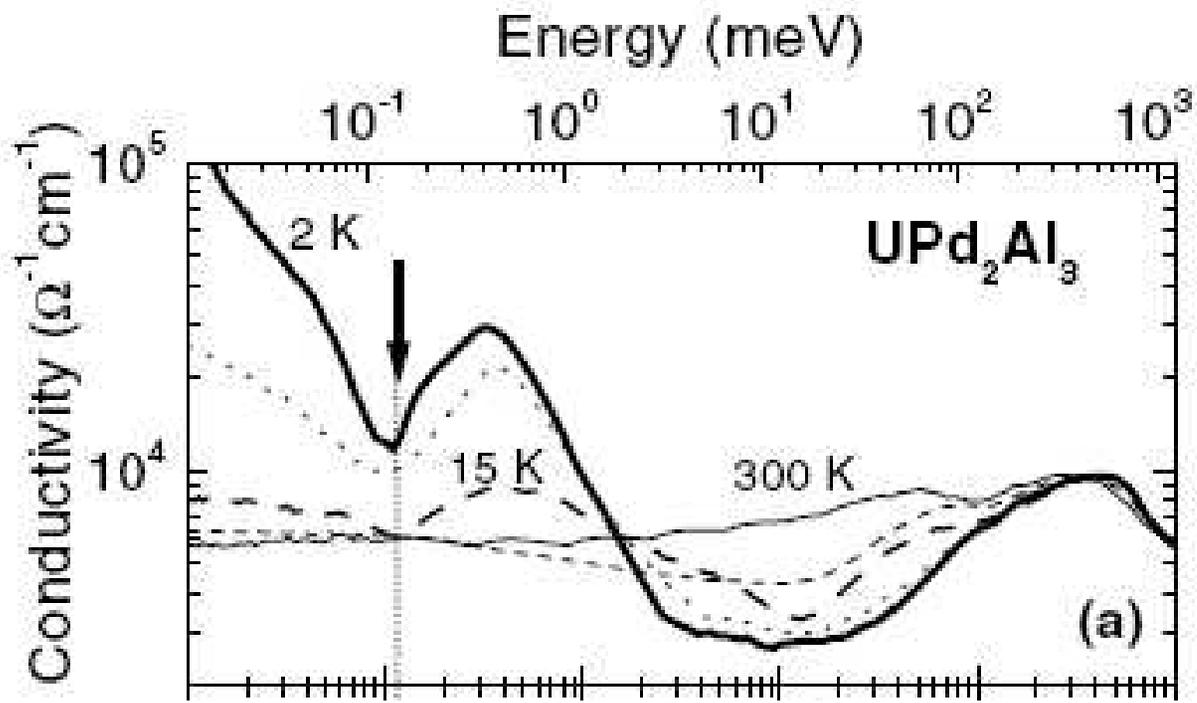
\begin{figure}[b]
\caption{The optical conductivity of a heavy fermion system
UPd$_2$Al$_{3}$ \cite{dressel02prl,dressel02prb}. Above coherence
temperature T$^* \approx$\,50\,K, the spectrum displays
conventional behaviour in the form of a broad Drude mode. Below
T$^* \approx$\,50\,K two distinctive feature develop in the
spectra: Drude-like mode at zero frequency and gap-like
excitations with an onset at around 50\,meV. In addition, below
AFM ordering temperature T$_N$=14\,K a new finite frequency mode
develops at around 5\,cm$^{-1}$, which causes further enhancement
of the effective mass.} \label{fig:hf}
\end{figure}

The electrodynamic response of heavy fermion metals (HF) has been
studied in a large number of systems \cite{degiorgi99}. Figure
\ref{fig:hf} displays optical conductivity of UPd$_2$Al$_3$
\cite{dressel02prl,dressel02prb}. Above the coherence temperature
T$^* \approx$\,50\,K $\sigma_1(\omega)$ is characterized by a
broad Drude-like peak at zero energy. Below T$^* \approx$\,50\,K,
but still above the AFM ordering temperature T$_N$=14\,K, two
notable features dominate the IR Spectrum of UPd$_2$Al$_3$: a very
narrow Drude-like mode and a gap like excitation, with an onset
around 50\,meV. One also observes a crossing of low- and high--T
$\sigma_1(\omega)$ spectra, with characteristic redistribution of
spectral weight. The electronic spectral weight removed from the
far-IR part of the spectrum is redistributed both to higher
frequencies (above the gap) and to a narrow Drude-like mode at
zero frequency.  The spectral weight in the Drude mode is smaller
but significant since it is ultimately responsible for the
metallic behavior of HF at low temperatures. This minuscule
spectral weight also indicates that the effective mass of the
quasiparticles is enhanced at low temperatures ($\omega_p^{*2}
\sim 1/m^*$).

The extended Drude model (Eqs.~\ref{eq:tau} and \ref{eq:mass}) has
been used frequently for the analysis of correlation effects in HF
\cite{webb86,bonn88,wachter94,donovan97,degiorgi99,dordevic01}.
The effective mass spectra (Eq.~\ref{eq:mass}) are especially
useful in that regard, as the quasiparticle effective mass m$^*$
can be extracted by extrapolating the $m^*(\omega)/m_b$ spectra
down to zero frequency. These so-called optical masses are usually
in good agreement with the values obtained using other techniques
(specific heat, de Hass--van Alphen, etc.)
\cite{bonn88,degiorgi99,dordevic01}. In UPd$_2$Al$_3$ the
effective mass is enhanced by a factor of  35 compared to the band
value $m_b$. Once the system crosses into the AFM ordered state,
below T$_N$=14\,K, a new mode develops around 5\,cm$^{-1}$. This
leads to further enhancement of the effective mass to about
50$m_b$, in good agreement with specific heat measurements.
Similar features were observed in another canonical HF UPt$_3$
\cite{dressel02prl,tran02}, which below 5\,K progressively
develops into a magnetically ordered state, with extremely small
local moments (0.02\,$\mu_B$) and without long range order.

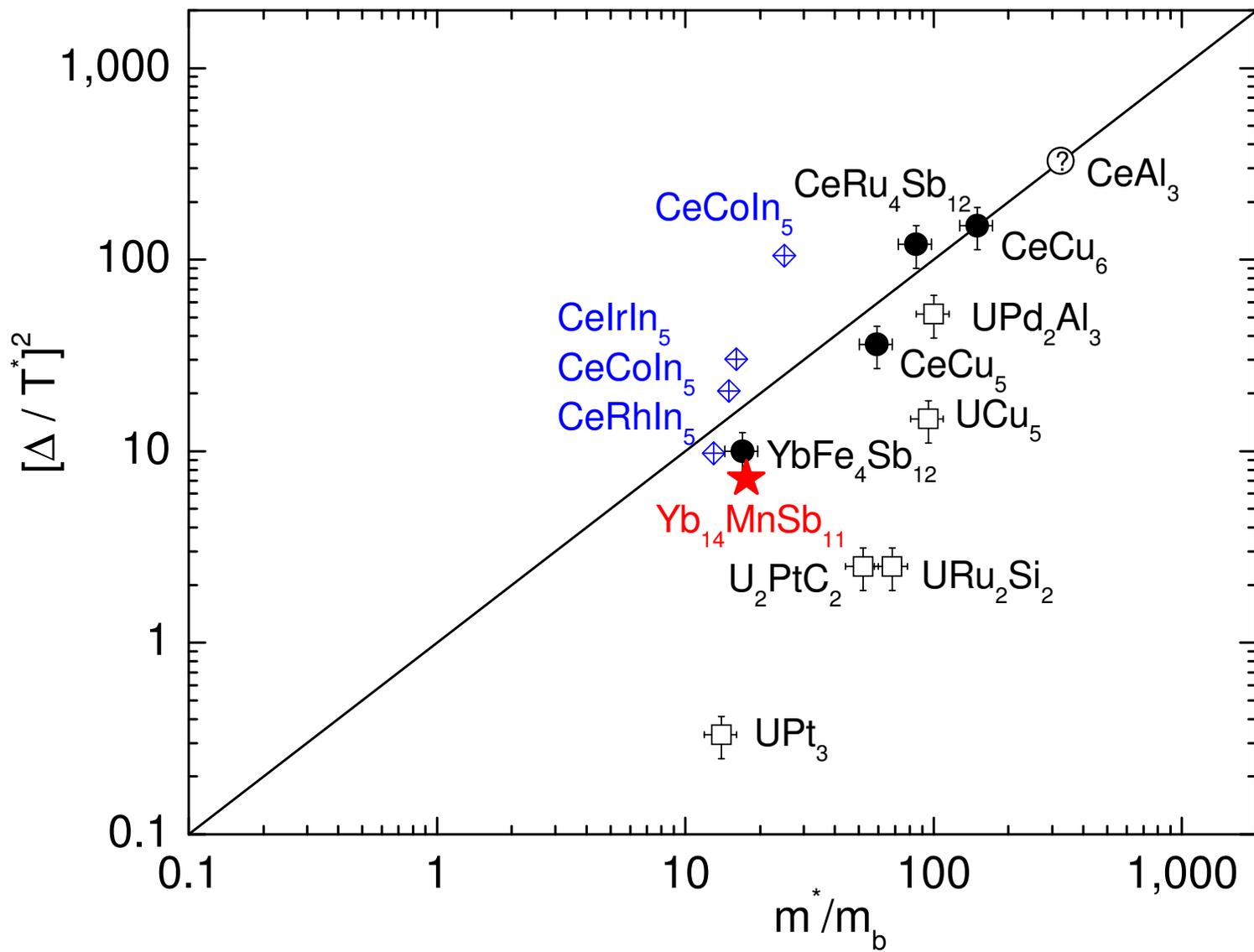
\begin{figure}[b]
\caption{Scaling between the quasiparticle effective mass
$m^*/m_b$ and the magnitude of hybridization gap $\Delta$ in a
number of heavy fermion systems \cite{dordevic01}. The line
represents $m^*/m_b=(\Delta/T^*)^2$ relation, where T$^*$ is the
coherence temperature. Notably, heavy fermion systems with
magnetic order do not fall on the scaling line. Similar scaling
between the optical gap and the effective mass has also been
discussed for CDW systems (Fig.~\ref{fig:cdw-scaling}).}
\label{fig:HFscaling}
\end{figure}

Hancock {\it et al.} studied the Kondo systems
YbIn$_{1-x}$Ag$_x$Cu$_4$ \cite{hancock04}, whose properties are
believed to be driven by similar physics as in HF metals. In
YbIn$_{1-x}$Ag$_x$Cu$_4$ silver doping $x$ can be continuously
changed over the whole range from 0 to 1. The physical properties
of the system also change from a moderately heavy fermion
YbAgCu$_4$ to a mixed-valence YbInCu$_4$. For all doping levels
the spectra reveal a finite frequency (2,000\,cm$^{-1}$) peak,
presumably due to excitations across the hybridization gap.
Hancock {\it et al.} showed that the energy of this mode scales
with the square root of the Kondo temperature and concluded that
the same underlying physics governs the thermodynamic properties
and the formation of this 2,000\,cm$^{-1}$ peak. Somewhat
different scaling has also been predicted \cite{millis87} and
experimentally observed in non-magnetic HF between the magnitude
of hybridization gap $\Delta$ and the quasiparticle effective mass
$m^*/m_b$ \cite{dordevic01}:

\begin{equation}
\frac{m^*}{m_b} \sim \Big( \frac{\Delta}{T^*} \Big)^2,
\label{eq:hf-scaling}
\end{equation}
where T$^*$ is coherence temperature (Fig.~\ref{fig:HFscaling}).
This scaling relation reflects the fact that in the
low-temperature coherent state of HF the intra--band response (as
characterized by $m^*/m_b$) and inter-band response (represented
by $\Delta$) are intimately related, at least in the simplest
version of the theory. Fig.~\ref{fig:HFscaling} also includes
recently studied d-electron system Yb$_{14}$MnSb$_{11}$, the first
case where heavy fermion behavior was discovered by IR
spectroscopy \cite{burch05}. We also note that this scaling
between the effective mass and the magnitude of the gap is similar
to the scaling in CDW systems Eq.~\ref{eq:cdw-scaling}, as
discussed in Section \ref{cdw} above. Similar scaling relations
are not surprising, knowing that the electrodynamic properties of
both systems are described by similar equations
\cite{sham97,portengen-thesis}.

IR spectroscopy has recently been used to study the $\alpha
\rightarrow \gamma$ transition in elemental Ce \cite{dirk01}. This
rare-earth metal with f-electrons undergoes a phase transition
from high temperature $\gamma$ to low temperature $\alpha$ phase.
The nature of this phase transition has been a matter of debate.
Reported infrared spectra strongly resemble IR spectra of
canonical HF metals. The high-T ($\gamma$) phase is characterized
by a broad and featureless spectrum; in the low-T ($\alpha$) phase
a narrow Drude-like mode and a peak at around 1\,eV  develop.
LDA+DMFT calculations by Haule {\it et al.} \cite{kotliar05} have
been able to reproduce both these features. They also revealed
that the 1\,eV peak is indeed due to excitations across
hybridization gap, formed by mixing of $f$-electrons with
conduction electrons.

Holden {\it et al.} \cite{reedyk03} studied the electrodynamic
response of a heavy fermion superconductor UBe$_{13}$, both in the
normal and superconducting state. At high temperatures a
conventional Drude-like behavior was observed, with essentially
frequency independent scattering rate. In the coherent state, but
still above T$_c$=0.9\,K, Holden {\it et al.} found a development
of a vary narrow zero-energy mode and a finite frequency peak,
presumably due to excitations across a hybridization gap. Finally,
in the superconductng state, below T$_c$, optical spectra revealed
marked changes in the low frequency region. In particular,
narrowing of the Drude peak and further suppression of optical
conductivity $\sigma_1(\omega)$ compared to normal state values
were observed. This behavior is not expected for a dirty limit
superconductors and has led Holden {\it et al.} to suggest that
UBe$_{13}$ is a clean limit superconductor.


\section{Dimensional crossover in organic and inorganic systems}
\label{1d}

Many familiar concepts of condensed matter physics are revised in
one dimension (1D) \cite{Dressel-review}. For example, the
conventional quasiparticle description breaks down in 1D solids
and the spin-charge separation paradigm needs to be invoked to
understand excitations. From a theoretical standpoint, the 1D
conductors are interesting in the context of Tomonaga-Luttinger
(TL) liquid description predicting unconventional power laws in
transport properties and also numerous ground states in systems
of coupled 1D chains \cite{giamarchi97,Tsvelik01}. An arrangement
of coupled 1D conductors is envisioned as a paradigm to explain
unconventional properties at higher dimensions specifically in the
context to the problem of high-$T_{c}$ superconductivity
\cite{clark94}. Signatures of 1D transport are being explored in a
wide variety of systems including but not limited to carbon
nano-tubes, conducting molecules, and semiconducting quantum wires, as
well as stripes in high-$T_{c}$ superconductors and quantum Hall
structures. In this section we analyze common characteristics of
the electromagnetic response of of quasi-1D organic conductors and
also of Cu-O chains in high-$T_c$ superconductor of YBCO family.

Infrared conductivity of quasi-1D conductors is highly
unconventional
\cite{tanner76,tanner83,timusk84,timusk85,cao96,Vescoli98}.
Several groups have investigated the response of linear chains of
tetramethyltetrathiafulvalene (TMTTF)$_2$X and
tetramethyltetraselenafulvalene (TMTSF)$_2$X, with different
interchain counterions X (such as X= AsF$_6$, PF$_6$, ClO$_4$,
Br). Charge transfer of one electron from every two (TMTSF)$_2$X
molecules leads to a quarter-filled (or half-filled bands due to
dimerization) hole band. In a purely 1D situation this would lead
to an insulating (Mott) state. However, Bechgaard salts are better
characterized as being only quasi-one-dimensional with finite
inter-chain hopping integrals which are distinct in the two
transverse directions. Both (TMTTF)$_2$X and (TMTSF)$_2$X families
of materials as well as other classes of organic quasi-1D
conductors typically show high dc conductivity along the direction
of the linear chains. One is therefore led to expect a metallic
reflectivity in the far infrared. Contrary to this expectation,
experiments typically show lower conductivity. The discrepancy
between DC and IR results can be resolved assuming the existence
of a narrow mode at $\omega=0$ with the width smaller than the
low-$\omega$ cut-off of the optical measurements. A direct
observation of a mode like this would provide strong argument in
favor of collective transport along the conducting chain. Cao {\it
et al.} analyzed the limitations on the spectral weight of this
mode and the width of the resonance based on the existing
reflectance data \cite{cao96}.

Optical conductivity data are capable of resolving the so-called
"dimensionality crossover" in the response of the linear chain
compounds \cite{giamarchi97} as first demonstrated by Vescoli {\it
et al.} \cite{Vescoli98}. In a purely 1D situation linear chain
systems are expected to be in a (Mott) insulating state because of
1/2 or 1/4 filled bands. However this is no longer the case if
inter-chain hopping integrals are finite, as is the case in
Bechgaard salts. Nevertheless, at high temperatures or frequencies
the interchain coupling is diminished thus enabling experimental
access to the 1D physics even in a system of weakly coupled
chains. The studies of the temperature dependence of the transport
properties or of the frequency dependence of the optical
conductivity then allow one to explore a crossover from the
regimes where 1D physics dominates (high-T, high-$\omega$) to the
regime where interchain coupling dominates (low-T, low-$\omega$).
The low energy response of (TMTSF)$_2$X compounds is governed by a
correlation gap leading to a sharp increase of the conductivity up
to a peak that signifies the magnitude of the gap $E_g$. This
behavior is followed by the power law dependence of the
conductivity at higher frequencies predicted theoretically
\cite{giamarchi97,giamarchi04}:

\begin{equation}
\sigma(\omega) \propto \omega^{4n^2} K_\rho^{-5}
\end{equation}
where $K_{\rho}$ is the TL-liquid exponent and n is the
commensurability. The authors have been able to show that the
experimental data indeed follow the power law dependence signaling
a crossover to 1D transport in mid-infrared frequencies.

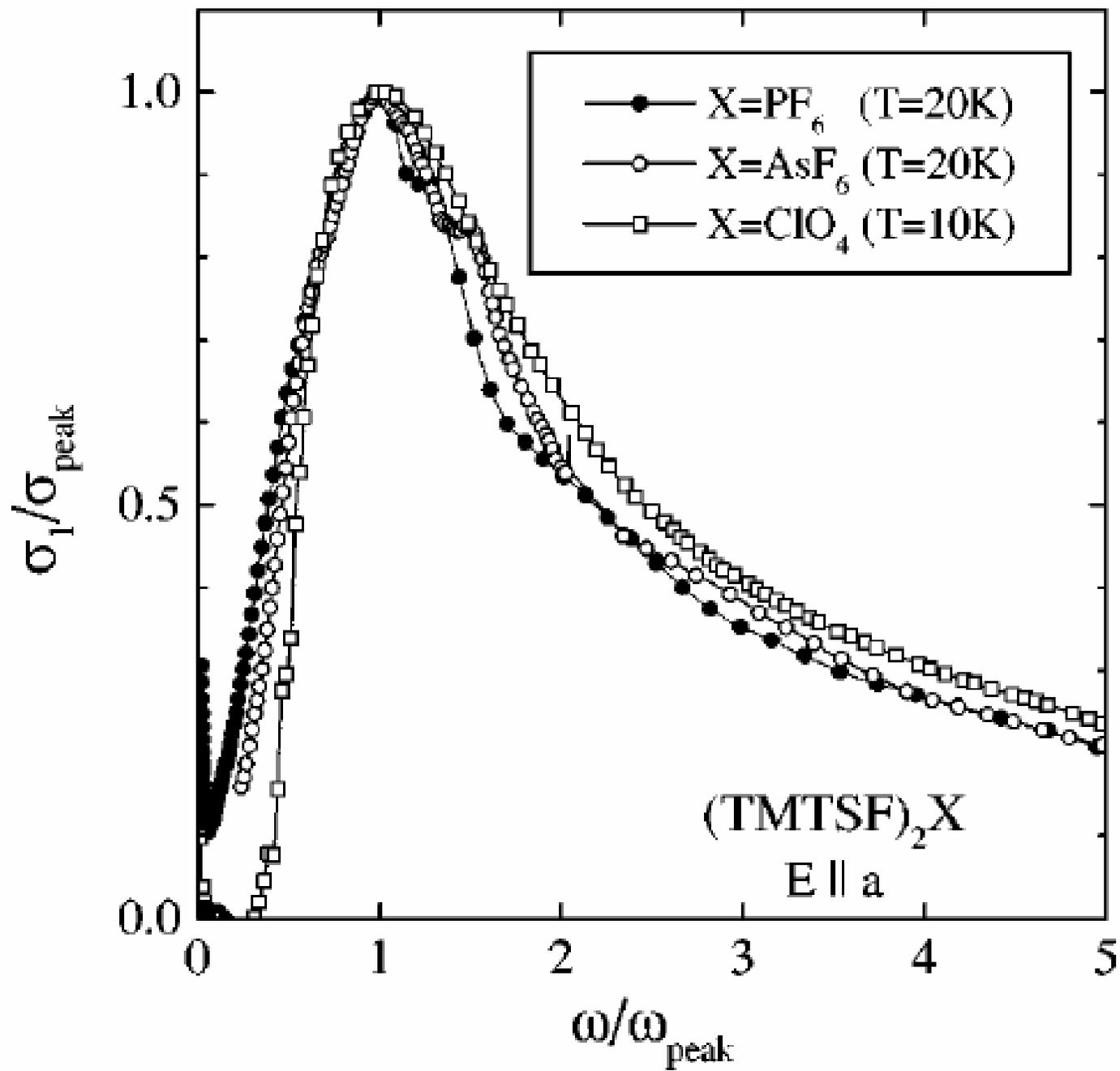
\begin{figure}[b]
\caption{The frequency-dependent conductivities of (TMTSF)$_2$X
(X=PF$_6$, AsF$_6$, and ClO$_4$) obtained in the polarization
along the conducting linear chains. Both axes have been normalized
to the peak of the finite energy mode near 200 cm$^{-1}$. The
universal behavior of the real part of the conductivity
$\sigma_1(\omega)$. From Schwartz {\it et al.} \cite{Schwartz98}.}
\label{fig:1dorganic}
\end{figure}

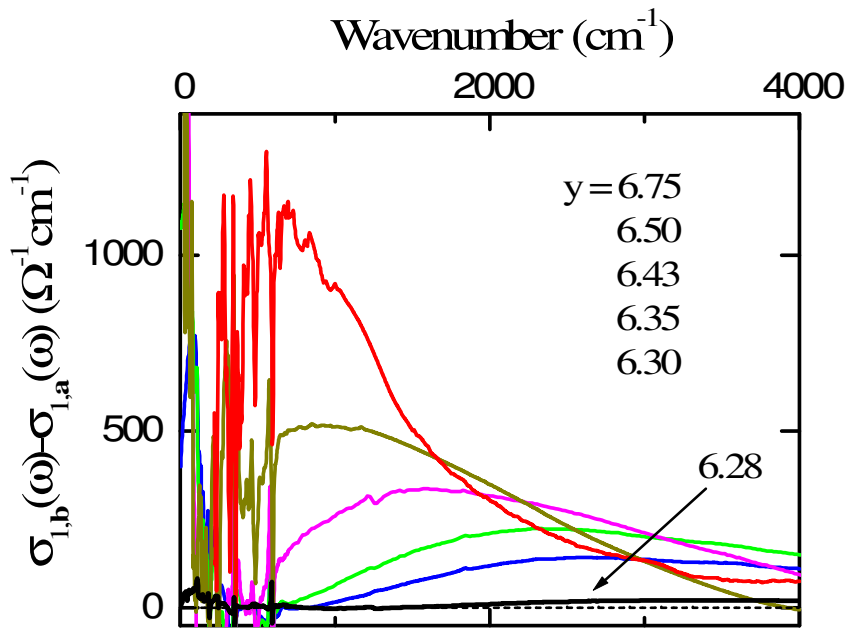
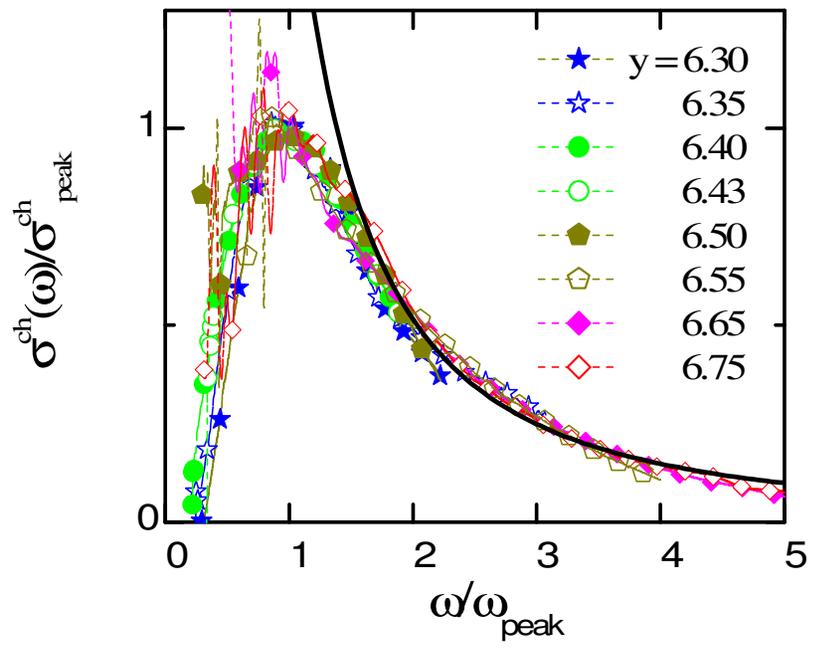
\begin{figure}[b]
\caption{Spectra of $\protect\sigma ^{ch}(\protect\omega
)=\protect \sigma _{1,b}(\protect\omega )-\protect\sigma
_{1,a}(\protect\omega )$ at lowest temperatures for a series of
YBCO crystals. For $y=6.75$, the coherent mode is not visible
because its weight is transferred to superconducting
$\protect\delta $-peak at $\protect\omega =0$. (b) $\protect
\sigma ^{ch}(\protect\omega )/\protect\sigma _{\text{peak}}^{ch}$
with $ \protect\omega /\protect\omega _{\text{peak}}$ as the
abscissa. For clarity, the sharp phonon structures are removed
from the data in the bottom panel. The solid line represents
$\protect\omega ^{-\protect\alpha }$-dependence with
$\protect\alpha =1.6$.}%
\label{fig:1dybco}
\end{figure}

To the best of our knowledge, similar power law scaling is not
observed in any inorganic quasi-1D compounds. An interesting
exception is the response of the Cu-O chains in
PrBa$_2$Cu$_4$O$_8$ discovered by Takenaka {\it et al.}
\cite{takenaka00}. In this family of high-T$_c$ cuprates as well
as in closely related YBa$_2$Cu$_3$O$_y$ compounds CuO$_2$
bilayers are separated with 1-dimensional Cu-O chains directed
along the b-axis. Assuming that the $b$-axis conductivity can be
described as a two-channel process one can isolate the chain
response as $\sigma^{ch} =\sigma^b-\sigma^a$ where $\sigma^b$ and
$\sigma^a$ are the conductivities probed along and across the
chain directions respectively. This analysis uncovered the power
law behavior of the chain conductivity $\sigma^{ch}$ consistent
with the response of  Bechgaard saults. Lee {\it et al.}
\cite{lee-prl05} examined scaling dependence of the $\sigma^{ch}$
spectra for a variety of compounds of YBa$_2$Cu$_3$O$_y$ series
with y=6.28-6.75. Despite the significant doping dependence of the
spectra, they all showed a universal behavior when the scaling
protocol of Schwartz {\it et al.} \cite{Schwartz98} was applied
(Fig.~\ref{fig:1dorganic}). The power law reported by Lee {\it et
al.} with $\alpha =1.6$ in Fig.~\ref{fig:1dybco} is distinct from
the $\omega ^{-3}$ response expected for a band insulator
\cite{giamarchi97}, but is close to $\alpha =1.3$ seen in 1D
Bechgaard salts \cite{Vescoli98,Schwartz98}. The range of values
of the correlation gap $E_{g}$ in YBa$_2$Cu$_3$O$_y$ is comparable
to that of the Bechgaard salts as well
\cite{Vescoli98,Schwartz98}.

Even though the mid-IR response of 1D Cu-O chains in cuprates and
that of the organic linear chain compounds uncover common trends
the low-frequency behavior of these two classes of 1D conductors
is radically different. Specifically, the low-energy collective
mode in Bechgaard salt is responsible for less than 1 $\%$ of the
total spectral weight of the infrared conductivity. In the YBCO
system this contribution is as high as 50 $\%$. Lee {\it et al.}
proposed that this discrepancy may originate from strong coupling
of the Cu-O chains to the conducting CuO$_2$ planes
\cite{lee-prl05}. This coupling appears to be enhanced with the
increasing doping. The frequency dependence of the collective mode
is sensitive to the amount of doping and impurities. At relatively
low dopings when chain segments are disordered the collective mode
has the form of a finite frequency resonance
\cite{lee-prl05,homes99}. Once the chain fragment length exceeds
the critical value and the separation between these fragments is
reduced, the $\sigma ^{ch}(\omega )$ spectra reveal the Drude-like
metallic behavior which would be impossible in a system of
isolated disordered chains. Finally we also note that no scaling
has been observed in the in-plane conductivity data of high-T$_c$
cuprates in the stripe ordered state discussed in Section
\ref{cdw}. On the other hand, IR measurements of the out-of-plane
penetration depth in YBCO have indicated the possibility of a
dimensional cross--over from 2D to 3D
\cite{homes04nature,homes05}.


\section{Summary }
\label{summary}

Infrared spectroscopy has been instrumental in elucidating a
number of interesting effects attributable to strong correlations
in solids. In these systems competing interaction often lead to a
formation of the energy gap or pseudogap that dominates the
electromagnetic response at low energies. Two classes of
correlated systems: density wave compounds and heavy fermion
conductors reveal correlations between the magnitude of the energy
gap and enhancement of quasiparticle effective mass. Despite
fundamental differences in the microscopic origins of the gapped
state the theoretical description of the electromagnetic response
is quite common.  One can identify similarities between the
spectroscopic fingerprints of pseudogaps in correlated transition
metal oxides with the response of the density wave or
hybridization gap materials.

Another common attribute of many correlated systems is strong
coupling of quiasiparticles to collective excitations. Infrared
optics uncovered many interesting characteristics of this
coupling. An inversion analysis that exemplified here with the
high-T$_c$ results is capable of yielding the details of the
relevant spectral function. It will be instructive to apply this
analysis will be applied to other materials as well. An important
virtue of infrared methods in this regard is that this technique
is a bulk probe. Other experimental methods capable of
investigating strong coupling effects (tunneling, ARPES) are
surface sensitive techniques.


\begin{acknowledgement}
We thank our collaborators T.~Timusk, C.C.~Homes, J.P.~Carbotte,
A.~Chubukov, L.~Degiorgi, W.J.~Padilla, K.S.~Burch and A.~LaForge
for useful discussions and critical comments on the manuscript.
\end{acknowledgement}


\begin{figure}[b]
\caption{The optical conductivity of several doped transition
metal insulators: La$_{1-x}$Sr$_x$TiO$_3$ \cite{fujishima-92},
La$_{1-x}$Sr$_x$VO$_3$ \cite{inaba95}, La$_{1-x}$Sr$_x$MnO$_3$
\cite{okimoto97}, La$_{1-x}$Sr$_x$CoO$_3$ \cite{tokura98},
La$_{2-x}$Sr$_x$NiO$_4$ \cite{ido91}, and La$_{2-x}$Sr$_x$CuO$_4$
\cite{uchida91}. The panels are arranged according to the number
of d electrons of transition metals, starting with Ti with
electron configuration [Ar]3d$^2$4s$^2$, until Cu with
configuration [Ar]3d$^{10}$4s. In all systems as doping $x$
increases the gap is filled, as the spectral weight from the
region above the gap is transferred below the gap and into the
Drude-like mode.} \label{fig:doped-insulators}
\end{figure}

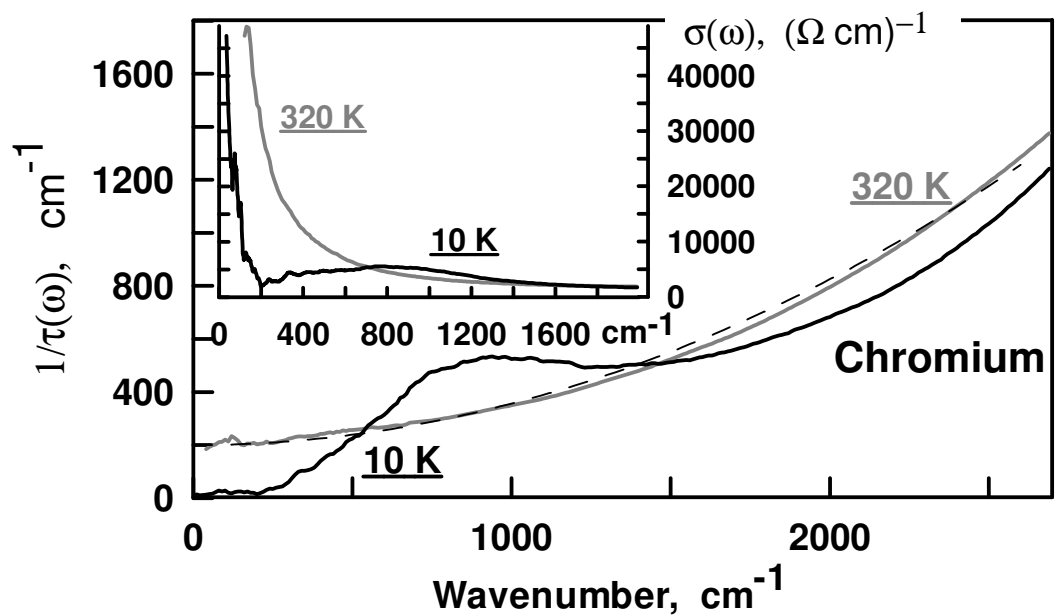
\begin{figure}[b]
\caption{The scattering rate 1/$\tau(\omega)$ and optical
conductivity $\sigma_1(\omega)$ (inset) of metallic chromium
\cite{basov02} at 10 and 312\,K. Chromium is a canonical example
of a 3D SDW system with T$_{SDW}$=\,312\,K. Above the transition
the optical constants display typical metallic behavior, with
a Drude-like mode in $\sigma_1(\omega)$ and $\omega^2$ dependence
of 1/$\tau(\omega)$ expected within Landau Fermi liquid theory. At
10\,K the spectra show deviations from this canonical behavior:
the conductivity reveals gap-like suppression between
200--700\,cm$^{-1}$ and narrowing of the zero-energy mode. This
results in suppression of scattering rate at low frequencies and
characteristic overshoot around 900\,cm$^{-1}$.} \label{fig:cr}
\end{figure}

\end{document}